\documentclass[preprint,12pt]{elsarticle}
\usepackage[utf8]{inputenc}
\usepackage[T1]{fontenc}
\usepackage{lmodern}
\usepackage{geometry}
\usepackage{amsmath,amssymb,amsthm}
\usepackage{graphicx}
\usepackage{booktabs}
\usepackage[dvipsnames]{xcolor}
\usepackage[colorlinks=true,linkcolor=blue,citecolor=blue,urlcolor=blue]{hyperref}

\theoremstyle{plain}
\newtheorem{proposition}{Proposition}
\newtheorem{corollary}{Corollary}

\newcommand{\Dr}{\mathring{\mathrm{D}}}
\newcommand{\omr}{\mathring{\omega}}
\newcommand{\dd}{\mathrm{d}}
\newcommand{\DD}{\mathrm{D}}
\newcommand{\II}{\mathrm{I}}
\newcommand{\vol}{v_{4}}
\newcommand{\K}{\,\mathrm{K}}

\begin{document}

\begin{frontmatter}

\title{Torsion talks only to spin: the CMB temperature as a test of energy exchange in Einstein--Cartan cosmology}

\author[uss]{Fernando Izaurieta\corref{cor1}}
\ead{fernando.izaurieta@uss.cl}
\author[pucv]{Samuel Lepe}
\ead{samuel.lepe@pucv.cl}
\author[uss]{Cristian Quinzacara}
\ead{cristian.quinzacara@uss.cl}

\cortext[cor1]{Corresponding author.}

\affiliation[uss]{organization={Departamento de Ciencias Exactas, Facultad de Ingeniería, Universidad San Sebastián},
            addressline={Lientur 1457},
            city={Concepción},
            postcode={4080871},
            country={Chile}}
\affiliation[pucv]{organization={Instituto de Física, Pontificia Universidad Católica de Valparaíso},
            addressline={Av.\ Brasil 2950},
            city={Valparaíso},
            postcode={2340025},
            country={Chile}}

\begin{abstract}
The cosmic microwave background temperature at $z=0.886$ is known to 2\% from molecular absorption towards PKS~1830$-$211; CO absorbers and Sunyaev--Zel'dovich clusters extend the measurement to $z\simeq3$. We use them on a question that cosmologies with torsion leave open. When the geometry exchanges energy with several matter components, which of them receives it? In Einstein--Cartan theory the Noether identities of each sector fix the answer. A sector's energy--momentum fails to be conserved only in proportion to its own spin, so minimally coupled light, which has none, is conserved whatever the torsion does. A metric-affine cosmology fitted to DESI DR2 distributes the exchange ``universally'' among all fluids, radiation included; the photons then gain almost a third of their energy. If the photons stay a blackbody, as FIRAS requires of any source set by local conditions, the model predicts a background colder than $T_{0}(1+z)$ by 7\% at $z\simeq0.9$ and by 8.5\% at $z\gtrsim3$, with no free parameter. PKS~1830$-$211 gives $5.08\pm0.10\K$ against a predicted $4.78\K$; with the CO absorbers it favours the standard scaling by $\Delta\chi^{2}=13$, and 18 Planck cluster bins give $\chi^{2}=50.8$ versus $15.5$. Under the same reading the primary anisotropies would change their spectral shape by amounts Planck excludes. If dark matter carries the spin instead, $H(z)$ moves by less than $10^{-4}$ below $z=3$ at the fitted parameters and $T(z)$ is standard again.
\end{abstract}

\begin{keyword}
torsion \sep Einstein--Cartan theory \sep interacting dark sector \sep cosmic microwave background temperature \sep quasar absorption lines \sep Noether identities
\end{keyword}

\end{frontmatter}

\section{Introduction}
\label{sec:intro}

In the direction of the quasar PKS~1830$-$211 there is a galaxy at redshift $z=0.886$ whose molecules absorb the quasar light. The populations of their rotational levels are in equilibrium with the cosmic microwave background (CMB), so the galaxy is a thermometer left there some seven billion years ago. Read with nine molecular species it marks $5.08\pm0.10\K$~\cite{Muller2013}; read with the HCN transitions alone, $5.13\pm0.06\K$~\cite{Kotani2025}. The standard prediction, $T_{0}(1+z)$ with $T_{0}=2.7255\K$~\cite{Fixsen2009}, is $5.14\K$. Further away, the rotational levels of CO and the fine structure of neutral carbon in quasar absorbers say the same thing out to $z\simeq3$: writing $T=T_{0}(1+z)^{1-\beta}$, they give $\beta=-0.015^{+0.030}_{-0.028}$~\cite{Klimenko2020}, improving on the $-0.007\pm0.027$ of the first CO survey~\cite{Noterdaeme2011}, and water absorption at $z=6.34$ brackets the standard value as well~\cite{Riechers2022}. Closer to home, the Sunyaev--Zel'dovich (SZ) effect in galaxy clusters gives $\beta=0.009\pm0.017$~\cite{Hurier2014}, $0.012\pm0.016$~\cite{Luzzi2015} and $-0.007\pm0.013$~\cite{deMartino2015}. The CMB photons have done nothing but cool with the expansion.

Those measurements are a serious test for any cosmology in which matter exchanges energy with the geometry, and cosmologies with torsion do exchange. In Einstein--Cartan (EC) theory and in its metric-affine extensions, matter with spin or hypermomentum does not conserve its energy--momentum separately~\cite{Hehl1976,HMMN1995,ObukhovPuetzfeld2014}. In a Friedmann universe with a single spinning fluid the form of the exchange has long been known, from the Weyssenhoff-fluid cosmologies~\cite{ObukhovKorotky1987,Gasperini1986} to the recent work of Kranas, Tsagas, Barrow and Iosifidis (KTBI)~\cite{KTBI2019,Iosifidis2020,AIJS2025}. With several fluids a question appears that the total balance equation cannot answer. That equation is like a family's bank statement: it records how much came in and how much went out, but not who spent it. Which fluid receives the energy?

The question has become practical. The DESI DR2 baryon acoustic oscillation data~\cite{DESIDR2} have revived the programme of interacting dark sectors, and torsion offers one of the few ways of deriving the interaction from a gravitational action instead of writing it by hand. Anagnostopoulos and Iosifidis state the allocation problem plainly in a recent paper that fits to DESI DR2 a metric-affine cosmology with a spinning hyperfluid: total conservation ``does not uniquely determine how the geometric source term is distributed among the individual fluids'', and they adopt as a modelling hypothesis a universal coupling in which every fluid, radiation included, exchanges energy with the geometry in proportion to its own $(1+3w_{i})\rho_{i}$~\cite{AI2026}. Andrei, Iosifidis, J\"arv and Saal (AIJS) do the opposite in a neighbouring model: radiation, matter and the cosmological constant obey the standard continuity equation (their Eq.~(15c)) and only a ``hyperdust'' exchanges energy with the geometry (their Eq.~(15d)), a dark matter which, in their words, they ``just assume'' to carry spin~\cite{AIJS2026}. In both cases the allocation is written down rather than derived. Interacting dark energy cosmologies face the same situation, and the lesson there is well known: the exchange term is usually written by hand, except when it is derived from an action, and then it is fixed~\cite{Wang2016,Amendola2000,Pourtsidou2013,Bohmer2015}.

We show that with torsion there is nothing to choose either, and that the answer is short: torsion talks only to spin. If each matter sector has its own action, the Noether identities of that action, taken by itself, fix how much energy and momentum the sector exchanges with the geometry, and the result is linear in the spin of that same sector (Proposition~\ref{prop:noether}). A spinless sector is conserved with the Levi-Civita derivative, for any torsion and any dynamics that produces it. The relevant example is light. With minimal coupling the electromagnetic field enters the action through $F=\dd A$ and the metric, never through the connection, so for light the torsion is just one more source of curvature of the pseudo-Riemannian background, and its temperature scales as $1+z$ with no correction at all. None of this is new as an identity: the conservation laws of matter with microstructure in Riemann--Cartan geometry have been known for decades~\cite{Hehl1976,HMMN1995,ObukhovPuetzfeld2014}, and that minimally coupled electromagnetism does not feel torsion or non-metricity was stated clearly by Hehl, Obukhov and Puetzfeld~\cite{HehlObukhov2001,PuetzfeldObukhov2008}. In practice the spin-fluid literature has often conserved radiation separately from the spinning fluid~\cite{Gasperini1986,BHL2008,Poplawski2010}, which is what the rule prescribes, and we ourselves once assigned the spin to dark matter without this justification~\cite{ILV2020}; what we have not found is anyone reading the identities as the rule that allocates the exchange among several fluids, or drawing an observational consequence from them.\footnote{Martins and Vieira used $T(z)$ to bound the steady-state torsion mode of KTBI~\cite{MartinsVieira2024}, by a different route that we discuss in Section~\ref{sec:thermometer}.}

The consequence is the point of this paper, and it is observable. Under the universal allocation of Ref.~\cite{AI2026} the CMB photons receive from the geometry, after recombination, almost a third of their present energy. An energy balance does not say what that does to the spectrum, and here the FIRAS measurement does most of the deciding. It sees the sum of everything injected since recombination, redshifted to today, and a source whose spectral shape is set by local conditions must then heat the blackbody as a blackbody: any other shape would leave a distortion thousands of times larger than the one FIRAS allows. So the blackbody gets hotter, and with the authors' own best-fit parameters the background should be $4.78\K$ at the redshift of PKS~1830$-$211 rather than $5.14\K$, and 8.5\% colder than $T_{0}(1+z)$ at every redshift above three. Above $z\simeq1.7$ the deficit does not depend on that argument. By then the model has injected too little energy to hide more than a fraction of it, unless it places the new photons exactly at the molecular transitions. A common rescaling of every photon energy, the only injection no thermometer would notice, is ruled out by the very spectral lines from which the model's redshifts were measured. Then both kinds of thermometer apply, and they agree. Molecules and atoms in absorbing galaxies, which sample the radiation where they sit, prefer the standard scaling by $\Delta\chi^{2}=13$; the SZ clusters, which read a distortion carried from the cluster to us, say so with more weight. (The same reading would change the spectral shape of the primary anisotropies by amounts that Planck's calibration excludes; we note it in Section~\ref{sec:thermometer}, and keep the thermometers as the test, the CO absorbers above all, which need only the background, today's temperature and the catalogued redshifts.) The measurement constrains the model, but it also does something more useful: it says which sector the geometry is not exchanging energy with, the photons, and so where the exchange has to go. Let a dark sector with spin carry it instead, and the expansion history is unchanged at the fitted parameters while $T(z)$ returns to normal.

The paper follows the order of the derivation. Section~\ref{sec:ec} writes the theory in differential forms and obtains its field equations. Section~\ref{sec:noether} derives the Noether identities of each sector and their three corollaries: spinless sectors are conserved, without spin there is no torsion, and the total law is the sum of the sector laws. Only then, in Section~\ref{sec:flrw}, do we impose the Friedmann symmetry and reduce everything to ordinary differential equations, among them the allocation law (Proposition~\ref{prop:allocation}), which shows that a universal allocation amounts to giving photons spin (Proposition~\ref{prop:universal}). Section~\ref{sec:thermometer} takes that consequence to the data, explaining on the way which thermometers can see it, and Section~\ref{sec:reassignment} shows that changing the recipient leaves the expansion history of Ref.~\cite{AI2026} untouched and restores the standard temperature. We close with the relation to previous work, including our own~\cite{nogo}, with what the rule says about reheating by torsion, and with the limits of the result (Section~\ref{sec:discussion}). The appendices carry the derivations, the kinetic argument behind the thermometers (\ref{app:kinetic}), a remark on the closure of the background system of Ref.~\cite{AI2026} (\ref{app:closure}) and the numerical details (\ref{app:numbers}). A reader who wants the observational argument first can go directly to Section~\ref{sec:thermometer}; it needs only Eq.~\eqref{eq:Qi} and the first corollary of Section~\ref{sec:noether}.

\section{Einstein--Cartan gravity in differential forms}
\label{sec:ec}

We work with a vierbein 1-form $e^{a}=e^{a}{}_{\mu}\dd x^{\mu}$ and an independent Lorentz connection 1-form $\omega^{ab}=-\omega^{ba}$, with $\eta_{ab}=\mathrm{diag}(-1,1,1,1)$, $\epsilon_{0123}=+1$, $\kappa=8\pi G$ and $c=1$. Torsion and curvature are
\begin{equation}
T^{a}=\DD e^{a}=\dd e^{a}+\omega^{a}{}_{b}\wedge e^{b},\qquad
R^{ab}=\dd\omega^{ab}+\omega^{a}{}_{c}\wedge\omega^{cb}.
\end{equation}
The Levi-Civita connection $\omr^{ab}$ is the unique torsion-free one, $\dd e^{a}+\omr^{a}{}_{b}\wedge e^{b}=0$, and the difference between the two connections is the contortion 1-form,
\begin{equation}
K^{ab}\equiv\omega^{ab}-\omr^{ab},\qquad T^{a}=K^{a}{}_{b}\wedge e^{b}.
\label{eq:contortion}
\end{equation}
We write $\Dr$ for the covariant derivative built with $\omr$, and $\star:\Omega^{p}\rightarrow\Omega^{4-p}$ for the Hodge dual. With it we define the contraction operator $\II_{a}:\Omega^{p}\rightarrow\Omega^{p-1}$, which in four dimensions and with our signature reads $\II_{a}\alpha=-\star(e_{a}\wedge\star\alpha)$ for any $p$-form $\alpha$, with $e_{a}=\eta_{ab}e^{b}$; it coincides with the contraction with the dual vector $E_{a}=e_{a}{}^{\mu}\partial_{\mu}$~\cite{Izaurieta:2019dix,Barrientos:2019msu}.

The volume form is $\vol=e^{0}\wedge e^{1}\wedge e^{2}\wedge e^{3}$, and we shall use the 3-forms $\star e_{a}=\II_{a}\vol$, which satisfy $e^{b}\wedge\star e_{a}=\delta^{b}_{a}\vol$.

The action is that of Einstein--Cartan gravity plus the matter sectors,
\begin{equation}
S=\frac{1}{4\kappa}\int\epsilon_{abcd}\,R^{ab}\wedge e^{c}\wedge e^{d}+\sum_{i}S_{i}\left[e,\omega,\psi_{i}\right],
\label{eq:action}
\end{equation}
where each sector $i$ has its own fields $\psi_{i}$ and interacts with the others only through the geometry.\footnote{The gravitational term corresponds to $\tfrac{1}{2\kappa}\int\sqrt{|g|}\,R(\Gamma)\,\dd^{4}x$ with the orientation $\dd t\wedge\dd x\wedge\dd y\wedge\dd z$. Two sectors that interact directly, such as electrons and photons before recombination, count as a single sector for what follows; since no minimally coupled field of the Standard Model carries vector spin~\cite{Hehl1976,Shapiro2002}, the visible sector as a whole receives nothing through the vector channel whether or not its components are counted separately.} The currents of each sector are the 3-forms that multiply the variations of the geometry once $\psi_{i}$ satisfies its equations of motion,
\begin{equation}
\delta S_{i}=\int\left(\delta e^{a}\wedge\tau^{(i)}_{a}+\frac{1}{2}\,\delta\omega^{ab}\wedge\sigma^{(i)}_{ab}\right).
\label{eq:currents}
\end{equation}
Here $\tau_{a}$ is the canonical energy--momentum 3-form, defined at fixed connection $\omega^{ab}$; for a spinning sector its components are not symmetric, by~\eqref{eq:lorentz}, and it coincides with the symmetric metric (Hilbert) stress--energy tensor only for a spinless sector. The 3-form $\sigma_{ab}=-\sigma_{ba}$ is the spin, or Lorentz hypermomentum, which in EC theory is the only part of the hypermomentum the connection responds to. Throughout the paper ``spin'' means this current, the response of the sector's action to the connection, and not the helicity of its quanta: a Maxwell field carries helicity one and, with $F=\dd A$, has $\sigma_{ab}=0$.\footnote{In components, $\tau_{a}=\tau_{a}{}^{b}\star e_{b}$ and $\sigma_{ab}=\sigma_{ab}{}^{c}\star e_{c}$. For a perfect fluid, $\tau_{0}{}^{0}=-\rho$ and $\tau_{j}{}^{k}=p\,\delta_{j}^{k}$. Defining $\tau_{a}$ at fixed connection rather than at fixed contortion is the natural choice of the first-order formalism and the one used in Refs.~\cite{HMMN1995,AIJS2025,AI2026}; the two definitions differ by terms linear in $\sigma$ and coincide for a spinless sector.}

Varying~\eqref{eq:action} with respect to $e^{d}$ and $\omega^{ab}$ gives the field equations (\ref{app:forms}),
\begin{align}
\frac{1}{2}\,\epsilon_{abcd}\,R^{ab}\wedge e^{c}&=\kappa\sum_{i}\tau^{(i)}_{d},
\label{eq:ec-metric}\\
\frac{1}{2}\,\epsilon_{abcd}\,T^{c}\wedge e^{d}&=-\frac{\kappa}{2}\sum_{i}\sigma^{(i)}_{ab}.
\label{eq:ec-connection}
\end{align}
The first is Einstein's equation written with the curvature of the full connection. The second is algebraic. It contains no derivatives of the torsion, so torsion does not propagate and is fixed point by point by the total spin. The map $T^{c}\mapsto\epsilon_{abcd}T^{c}\wedge e^{d}$ is invertible in four dimensions, so~\eqref{eq:ec-connection} can be read in both directions: where there is no spin there is no torsion.

\section{The Noether identity of each sector}
\label{sec:noether}

Each $S_{i}$ is, on its own, invariant under diffeomorphisms and under local Lorentz transformations, and that is all the following proposition needs.

\begin{proposition}[Sector-wise balance]
\label{prop:noether}
On the equations of motion of $\psi_{i}$, the currents of each sector satisfy
\begin{align}
	e_a \wedge \tau_b - e_b \wedge \tau_a &=\DD\sigma_{ab} \label{eq:lorentz}\\
\Dr\tau_{a}&=\frac{1}{2}\left(\II_{a}R^{bc}\right)\wedge\sigma_{bc}-\frac{1}{2}\left(\II_{a}K^{bc}\right)\DD\sigma_{bc}.
\label{eq:diffeo}
\end{align}
\end{proposition}
\begin{proof}
Lorentz invariance, with $\delta e^{a}=\lambda^{a}{}_{b}e^{b}$ and $\delta\omega^{ab}=-\DD\lambda^{ab}$, gives~\eqref{eq:lorentz} after one integration by parts. Diffeomorphism invariance, with $\delta e^{a}=\DD\xi^{a}+\II_{\xi}T^{a}$ and $\delta\omega^{ab}=\II_{\xi}R^{ab}$ (a covariant Lie derivative; the accompanying Lorentz transformation is already accounted for in~\eqref{eq:lorentz}), gives the classical identity of Hehl and collaborators~\cite{Hehl1976},
\begin{equation}
\DD\tau_{a}=\left(\II_{a}T^{b}\right)\wedge\tau_{b}+\frac{1}{2}\left(\II_{a}R^{bc}\right)\wedge\sigma_{bc}.
\label{eq:hehl}
\end{equation}
The details, signs and factors included, are in~\ref{app:forms}. The last term of~\eqref{eq:diffeo} vanishes in the Friedmann universe of Section~\ref{sec:flrw}, so nothing in what follows depends on it.
\end{proof}

Equation~\eqref{eq:diffeo} is the allocation rule, and it is worth reading term by term. On the left is the Levi-Civita divergence of the sector's energy--momentum, that is, what an observer who only knows the metric would call ``non-conservation''. On the right everything is linear in the spin of the same sector, $\sigma_{bc}$ and its derivative, multiplied by geometric quantities, the curvature and the contortion, which are common to all sectors. What a sector receives depends on its own spin alone.

For a spinning sector, what counts as its energy is a matter of convention. We defined $\tau_{a}$ at fixed connection. Defining it at fixed contortion instead, $\tau'_{a}$, shifts it by terms linear in the sector's own spin, and the same derivation gives
\begin{equation}
\Dr\tau'_{a}=\frac{1}{2}\left(\II_{a}\Dr K^{bc}\right)\wedge\sigma_{bc}-\frac{1}{2}\left(\II_{a}K^{bc}\right)\Dr\sigma_{bc},
\label{eq:diffeo-contortion}
\end{equation}
with the contortion where~\eqref{eq:diffeo} has the curvature: in that bookkeeping a sector exchanges energy only through the derivatives of the contortion and of its own spin (the last term, which vanishes in the Friedmann universe). The difference between the two laws is the coupling of the spin to the Riemannian curvature, which~\eqref{eq:diffeo} keeps as an exchange and~\eqref{eq:diffeo-contortion} absorbs into the energy--momentum, in the manner of Belinfante and Rosenfeld (\ref{app:forms} derives~\eqref{eq:diffeo-contortion}). The form of the rule survives the change of bookkeeping, the amounts assigned to spinning sectors do not, and we shall say so where a number depends on it. For a spinless sector the two definitions coincide, and the result about photons does not depend on the choice.

\begin{corollary}[Spinless sectors]
\label{cor:blind}
If $\sigma^{(i)}_{ab}=0$ then $\Dr\tau^{(i)}_{a}=0$: the sector is conserved with the Levi-Civita derivative, whatever the torsion and whatever the gravitational dynamics that produces it.
\end{corollary}

For a sector whose action does not depend on $\omega$ the corollary is immediate, and does not even need Proposition~\ref{prop:noether}: its action sees only the metric, and diffeomorphism invariance of an action that sees only the metric gives $\Dr\tau_{a}=0$, exactly as in general relativity. This is the case of light. With minimal coupling, $S_{\gamma}=-\tfrac{1}{2}\int F\wedge\star F$ with $F=\dd A$, and neither $F$ nor the Hodge dual knows about the connection~\cite{Hehl1976,HehlObukhov2001,Shapiro2002}. That is why the corollary does not depend on the geometry being Einstein--Cartan: in any metric-affine theory, with torsion and non-metricity, and with any gravitational action, an action that does not see the connection gives $\Dr\tau_{a}=0$. For the electromagnetic field, torsion is one more source of curvature of the pseudo-Riemannian background it lives in, through~\eqref{eq:ec-metric}, and nothing else. The same holds for a minimally coupled scalar and for a spinless perfect fluid.

Dirac fermions are different. With minimal coupling their spin tensor is totally antisymmetric, $\sigma_{abc}\propto\epsilon_{abcd}J_{5}^{d}$, with $J_{5}^{d}=\bar{\psi}\gamma^{d}\gamma^{5}\psi$ the axial current, and couples only to the axial part of the torsion~\cite{Hehl1976,Shapiro2002}. Isotropy removes the spatial part of the axial current but not its density $n_{5}=J_{5}^{0}$, which breaks parity rather than isotropy: a net chirality, in the visible or the dark sector, sources the axial function $Y$ of Section~\ref{sec:flrw} and never the vector one.\footnote{This is also what Ma, Mudrunka and Nakayama find when they couple a dynamical torsion to the Standard Model: the torsion--fermion couplings appear on their own and the couplings to gauge bosons do not, by gauge invariance~\cite{MMN2026}. On the quantum side, Capolupo and collaborators quantise Dirac fields on a Friedmann background with torsion and find a divergence of the fermionic vacuum energy--momentum built from the vacuum's own axial current contracted with the torsion~\cite{Capolupo2026}, which is the structure of~\eqref{eq:diffeo}.} Since~\eqref{eq:diffeo} is linear in $\sigma$ and the background geometry is common, the rule also holds for an ensemble average, with the average spin in place of $\sigma$. If the spin of a sector fluctuates about a zero mean, the torsion it sources fluctuates with it, and the product of the two in~\eqref{eq:diffeo} averages to a term quadratic in the spin. That is the contact term of the averaged Weyssenhoff fluid~\cite{HHK1974,Hehl1976,ObukhovKorotky1987,Poplawski2012}. For fluctuations uncorrelated between sectors it behaves as effective matter belonging to the same sector; correlated fluctuations between two spinning sectors would be an exchange between those two. For a sector whose spin current vanishes identically, like Maxwell's, there is nothing to average.

There are exceptions, and they are precisely what the data can detect. A non-minimal coupling of fermions makes the trace of the torsion couple to the vector current, that is, to particle number~\cite{FreidelMinicTakeuchi2005,MZK2013}, and then baryons have spin in the sense of~\eqref{eq:currents}. A direct coupling of the electromagnetic field to torsion, of the kind discussed since Hojman, Rosenbaum, Ryan and Shepley~\cite{Hojman1978}, or a gauge triad whose field strength contains the torsion~\cite{KasemKhalil2021}, or light propagating along the autoparallels of the connection with torsion~\cite{Bolejko2020}, all give light a spin. None of these cases contradicts Proposition~\ref{prop:noether}. They are sectors with spin, and~\eqref{eq:diffeo} says how much they receive.

\begin{corollary}[No spin, no torsion]
\label{cor:nosource}
In EC theory, if no sector has spin, Eq.~\eqref{eq:ec-connection} gives $T^{a}=0$. A torsion ansatz imposed by hand on a universe whose matter is minimally coupled and carries no net spin (no chiral asymmetry, in the case of fermions) does not solve the field equations, and the exchange law that such an ansatz forces on the matter does not follow from the Einstein--Cartan action.
\end{corollary}

\begin{corollary}[The sum]
\label{cor:sum}
Summing~\eqref{eq:diffeo} over sectors and using~\eqref{eq:ec-connection} gives the total balance law, which involves only the total torsion and the total energy--momentum. That law is a consequence of Proposition~\ref{prop:noether}, but does not imply it: any redistribution of the right-hand sides among the sectors that leaves the sum invariant satisfies it.
\end{corollary}

Equations~\eqref{eq:ec-connection} and~\eqref{eq:diffeo} together say what the total law could not: each sector contributes its spin to the torsion, and receives energy in proportion to that same contribution.

\section{The Friedmann universe}
\label{sec:flrw}

We now impose the symmetry. In a spatially flat Friedmann--Lema\^itre--Robertson--Walker (FLRW) universe with vierbein $e^{0}=\dd t$, $e^{j}=a(t)\,\dd x^{j}$, the torsion compatible with homogeneity and isotropy has two functions~\cite{Tsamparlis1979}, a vector one, $X(t)$, and an axial one, $Y(t)$,
\begin{equation}
T^{0}=0,\qquad T^{j}=-X\,e^{0}\wedge e^{j}-Y\,\epsilon^{j}{}_{kl}\,e^{k}\wedge e^{l},
\label{eq:torsion-flrw}
\end{equation}
in the notation of Refs.~\cite{Cid2018,ILV2020}.\footnote{In the notation of KTBI~\cite{KTBI2019}, $X=2\phi$; in that of AIJS~\cite{AIJS2025} and of Ref.~\cite{AI2026}, $X=-\kappa\sigma/2$. Spatial curvature $k$ adds $+3k/a^{2}$ to the left-hand side of~\eqref{eq:friedmann} and $+k/a^{2}$ to that of~\eqref{eq:acceleration}, and changes nothing in what follows.} In this section and the next, $j,k,l$ are spatial frame indices and $i$ labels matter sectors. With $H=\dot{a}/a$ and the combination
\begin{equation}
Z\equiv H+X,
\end{equation}
which we call the \emph{Hubble rate of the connection}, the connection and its curvature are
\begin{align}
\omega^{0j}&=Z\,e^{j},\qquad\qquad\qquad\ \ \omega^{jk}=Y\,\epsilon^{jk}{}_{l}\,e^{l},
\label{eq:omega-flrw}\\
R^{0j}&=\frac{1}{a}\,\frac{\dd(aZ)}{\dd t}\,e^{0}\wedge e^{j}-ZY\,\epsilon^{j}{}_{kl}\,e^{k}\wedge e^{l},\qquad
R^{jk}=\left(Z^{2}-Y^{2}\right)e^{j}\wedge e^{k}+\frac{1}{a}\,\frac{\dd(aY)}{\dd t}\,\epsilon^{jk}{}_{l}\,e^{0}\wedge e^{l}.
\label{eq:R-flrw}
\end{align}
The Levi-Civita connection is~\eqref{eq:omega-flrw} with $Z\rightarrow H$ and $Y\rightarrow0$. In words: the vector torsion replaces $H$ by $Z$ in the boost part of the connection, and the axial torsion adds a rotation. The currents compatible with isotropy are, for each sector,
\begin{equation}
\tau^{(i)}_{0}=-\rho_{i}\star e_{0},\qquad \tau^{(i)}_{j}=p_{i}\star e_{j},\qquad
\sigma^{(i)}_{0j}=\sigma_{i}\star e_{j},\qquad \sigma^{(i)}_{jk}=\zeta_{i}\,\epsilon_{jk}{}^{l}\star e_{l},
\label{eq:currents-flrw}
\end{equation}
with four functions of time: density, pressure, vector spin $\sigma_{i}$ and axial spin $\zeta_{i}$. A spinless sector has $\sigma_{i}=\zeta_{i}=0$.

The field equations reduce by direct substitution (\ref{app:flrw}). The $d=0$ component of~\eqref{eq:ec-metric} and any one of its spatial components give
\begin{align}
3\left(Z^{2}-Y^{2}\right)&=\kappa\rho,
\label{eq:friedmann}\\
2\left(\dot{Z}+HZ\right)+Z^{2}-Y^{2}&=-\kappa p,
\label{eq:acceleration}
\end{align}
with $\rho=\sum_{i}\rho_{i}$ and $p=\sum_{i}p_{i}$, while the $(0j)$ and $(jk)$ components of~\eqref{eq:ec-connection} give the map between spin and torsion,
\begin{equation}
X=-\frac{\kappa}{2}\sum_{i}\sigma_{i},\qquad Y=\frac{\kappa}{2}\sum_{i}\zeta_{i},
\label{eq:map}
\end{equation}
which is that of AIJS~\cite{AIJS2025}, up to the conventional sign of the axial part. Subtracting one third of~\eqref{eq:friedmann} from~\eqref{eq:acceleration} cancels $Y^{2}$ and the spatial curvature, and what is left is the Raychaudhuri equation of the connection~\cite{PTB2017,ITP2018},
\begin{equation}
\frac{\dd(aZ)}{\dd t}=-\frac{\kappa a}{6}\left(\rho+3p\right),
\label{eq:raych}
\end{equation}
which by~\eqref{eq:R-flrw} is also the coefficient of $e^{0}\wedge e^{j}$ in $R^{0j}$.

The reduction of Proposition~\ref{prop:noether} is just as direct. In FLRW, isotropy makes $\DD\sigma_{ab}=0$ for the currents~\eqref{eq:currents-flrw}, so~\eqref{eq:lorentz} says that $\tau_{a}$ is symmetric, as it already is in~\eqref{eq:currents-flrw}, and the last term of~\eqref{eq:diffeo} drops out. The component of~\eqref{eq:diffeo} with frame index $a=0$ has on the left $\Dr\tau_{0}=-[\dot{\rho}_{i}+3H(\rho_{i}+p_{i})]\,\vol$, and on the right the contractions $\II_{0}R^{0j}=a^{-1}\,\dd(aZ)/\dd t\;e^{j}$ and $\II_{0}R^{jk}=a^{-1}\,\dd(aY)/\dd t\;\epsilon^{jk}{}_{l}e^{l}$ wedged with the spin 3-forms:
\begin{proposition}[Allocation in FLRW]
\label{prop:allocation}
For each sector,
\begin{equation}
\dot{\rho}_{i}+3H\left(\rho_{i}+p_{i}\right)=Q_{i}\equiv-\frac{3}{a}\left[\sigma_{i}\,\frac{\dd(aZ)}{\dd t}+\zeta_{i}\,\frac{\dd(aY)}{\dd t}\right]
=\frac{\kappa}{2}\,\sigma_{i}\left(\rho+3p\right)-3\,\zeta_{i}\left(\dot{Y}+HY\right),
\label{eq:Qi}
\end{equation}
where the second form uses~\eqref{eq:raych}.
\end{proposition}

The energy a sector receives is the work done by the curvature of the connection on its spin, $\tfrac{1}{2}(\II_{0}R^{bc})\wedge\sigma_{bc}$. Equation~\eqref{eq:Qi} is a balance, not a conservation law, and it becomes one exactly when the spin vanishes: a sector with $\sigma_{i}=\zeta_{i}=0$ receives nothing, however large $X$ and $Y$ are.\footnote{A spinning sector can also be conserved by accident, when its two channels cancel, $\tfrac{\kappa}{2}\sigma_{i}(\rho+3p)=3\zeta_{i}(\dot{Y}+HY)$. Nothing protects that cancellation, and we do not consider it further.} The vector channel is sourced by the total $\rho+3p$, the same combination that drives the Raychaudhuri equation~\eqref{eq:raych}; the axial channel by the variation of $aY$. Summing over sectors, with~\eqref{eq:map} and~\eqref{eq:raych}, one obtains $\sum_{i}Q_{i}=-X(\rho+3p)-(6/\kappa)Y(\dot{Y}+HY)$, the total law of Refs.~\cite{KTBI2019,ILV2020}; with a single sector, the vector part of~\eqref{eq:Qi} is the law of KTBI~\cite{KTBI2019}, of Iosifidis~\cite{Iosifidis2020} and of AIJS~\cite{AIJS2025}. The ambiguity pointed out in Ref.~\cite{AI2026} is visible here too: any redistribution $Q_{i}\rightarrow Q_{i}+G_{i}(t)$ with $\sum_{i}G_{i}=0$ leaves the total law untouched, and the sum cannot tell one from another. Proposition~\ref{prop:allocation} removes the ambiguity, because it fixes each $Q_{i}$ and not only the sum.

\begin{proposition}[Universal allocation]
\label{prop:universal}
The prescription
\begin{equation}
\dot{\rho}_{i}+3H\left(1+w_{i}\right)\rho_{i}=\frac{\kappa\sigma}{2}\left(1+3w_{i}\right)\rho_{i}\qquad\text{for all }i,
\label{eq:universal}
\end{equation}
with a single $\sigma$ and $Y=0$, coincides with~\eqref{eq:Qi} if and only if each sector carries the vector spin $\sigma_{i}=\sigma\,(1+3w_{i})\rho_{i}/(\rho+3p)$, wherever $\rho+3p\neq0$. In particular it requires radiation to carry $\sigma_{\gamma}=2\sigma\rho_{\gamma}/(\rho+3p)\neq0$.
\end{proposition}
\begin{proof}
Equate the vector part of~\eqref{eq:Qi} with the right-hand side of~\eqref{eq:universal}; the $\sigma_{i}$ so obtained add up to $\sigma$.
\end{proof}

The identification is a bookkeeping rule rather than a physical assignment. It ties the spin of a sector to its equation of state: a component with $w=-1/3$ would carry no spin, regardless of what it is made of, two dust components would share it in proportion to their densities, and the spin of each sector would change whenever its $w$ does. We know of no action that distributes spin that way, and it is hard to see how one could: spin is a property of what a sector is made of, not of its pressure.\footnote{Not even a cosmological constant would escape the prescription. With $w=-1$, Eq.~\eqref{eq:universal} assigns it $Q_{\Lambda}=-\kappa\sigma\rho_{\Lambda}$, so that $\rho_{\Lambda}\propto\exp(-\kappa\int\sigma\,\dd t)$; but $\Lambda$ has no spin, and Corollary~\ref{cor:blind} conserves it identically, as it must, since $\Lambda g_{\mu\nu}$ is covariantly constant for the Levi-Civita connection. AIJS keep their $\Lambda$ constant for this reason~\cite{AIJS2026}.}

If the radiation is the cosmic photon background, Proposition~\ref{prop:universal} and Corollary~\ref{cor:blind} contradict each other. Anagnostopoulos and Iosifidis present their construction as one in which the interactions appear ``while the Standard Model of Particle physics remains unmodified'', and they exclude direct energy exchange between the fluids precisely ``in order to maintain the integrity of the Standard Model of particle physics''~\cite{AI2026}. Read with Proposition~\ref{prop:noether}, their allocation nonetheless assigns the photons an exchange with the geometry. One could object that their hyperfluid is specified by its currents rather than by an action (variational hyperfluids exist~\cite{ObukhovTresguerres1993}, but none is invoked there), so that Proposition~\ref{prop:noether} has nothing to say about it. Whether or not that is so for the hyperfluid, it is not so for the photons: Maxwell's action exists, and it is the photon sector that Corollary~\ref{cor:blind} constrains. Either photons have a non-minimal coupling to the connection, of the kind listed in Section~\ref{sec:noether}, which should then appear in the action and would have consequences well beyond the background cosmology, or the universal allocation does not describe photons.

\section{The thermometer}
\label{sec:thermometer}

\subsection{Universal allocation and \texorpdfstring{$T(z)$}{T(z)}}

Ref.~\cite{AI2026} considers a perfect hyperfluid in the metric-affine framework whose hypermomentum is pure spin, with the shear and dilation parts set to zero and only the vector function $\sigma$ allowed by isotropy (their Eq.~(2.22) and Sec.~2.2).\footnote{Equation, table and section numbers of Ref.~\cite{AI2026} refer throughout to arXiv:2609.07978v1; \ref{app:closure} reproduces the equations we use.} Its background equations, their system~(2.24), coincide with~\eqref{eq:friedmann}, \eqref{eq:acceleration} and Corollary~\ref{cor:sum} with $Y=0$ and $X=-\kappa\sigma/2$:
\begin{equation}
3\left(H-\frac{\kappa\sigma}{2}\right)^{2}=\kappa\rho,\qquad
\dot{\rho}+3H\left(\rho+p\right)=\frac{\kappa\sigma}{2}\left(\rho+3p\right).
\label{eq:AIfried}
\end{equation}
Here $\rho$ is matter plus radiation. The dark energy of the model is geometric: the authors identify it with the spin contribution to the Friedmann equation, $\rho_{\rm DE}\equiv3(H^{2}-Z^{2})/\kappa=3H\sigma-3\kappa\sigma^{2}/4$ (their Eq.~(2.28)), and they write the universal coupling for matter, radiation and this effective component alike (their Eqs.~(2.33)--(2.35)). The model the authors prefer, ``polytropic spin'', has $\kappa\sigma/(2H_{0})=bE^{n}$, with $E=H/H_{0}$, and the first of Eqs.~\eqref{eq:AIfried} evaluated today fixes
\begin{equation}
b=1-\sqrt{\Omega_{m0}+\Omega_{r0}}.
\end{equation}
Under the universal allocation~\eqref{eq:universal} the two sectors obey $\dot{\rho}_{m}+3H\rho_{m}=\tfrac{\kappa\sigma}{2}\rho_{m}$ and $\dot{\rho}_{r}+4H\rho_{r}=\kappa\sigma\rho_{r}$, whose sum is the second of Eqs.~\eqref{eq:AIfried}. If the radiation keeps a Planck spectrum, a condition we return to in a moment, then $\rho_{r}\propto T^{4}$ and
\begin{equation}
\frac{\dd\ln T}{\dd\ln(1+z)}=1-\frac{\kappa\sigma}{4H}=1-\frac{b}{2}\,E^{\,n-1},
\qquad\text{i.e.}\qquad
T(z)=T_{0}\left(1+z\right)\exp\left[-\frac{\kappa}{4}\int_{0}^{z}\frac{\sigma\,\dd z'}{(1+z')H}\right]:
\label{eq:Tz}
\end{equation}
the standard scaling times the exponential of the energy the geometry has injected into the radiation since redshift $z$. The authors discuss this consequence for their simplest model, in which $\kappa\sigma/H$ is constant, identify the radiation with the CMB and note that the bounds on $\beta$ constrain it (their Eq.~(2.51))~\cite{AI2026}. For the preferred model they do not revisit it, and that is where it matters: today $E=1$, so the local slope is $\beta(0)=b/2$ for any $n$. With $\Omega_{m0}=0.380$, $\beta(0)=0.19$. The deviation does not depend on the parameter that controls the evolution of the spin, but on how much geometric dark energy the model needs today. No reasonable corner of its parameter space avoids it: bringing $\beta(0)$ down to $0.03$ would require $\Omega_{m0}>0.88$.

\begin{figure}[t]
\centering
\includegraphics[width=0.98\textwidth]{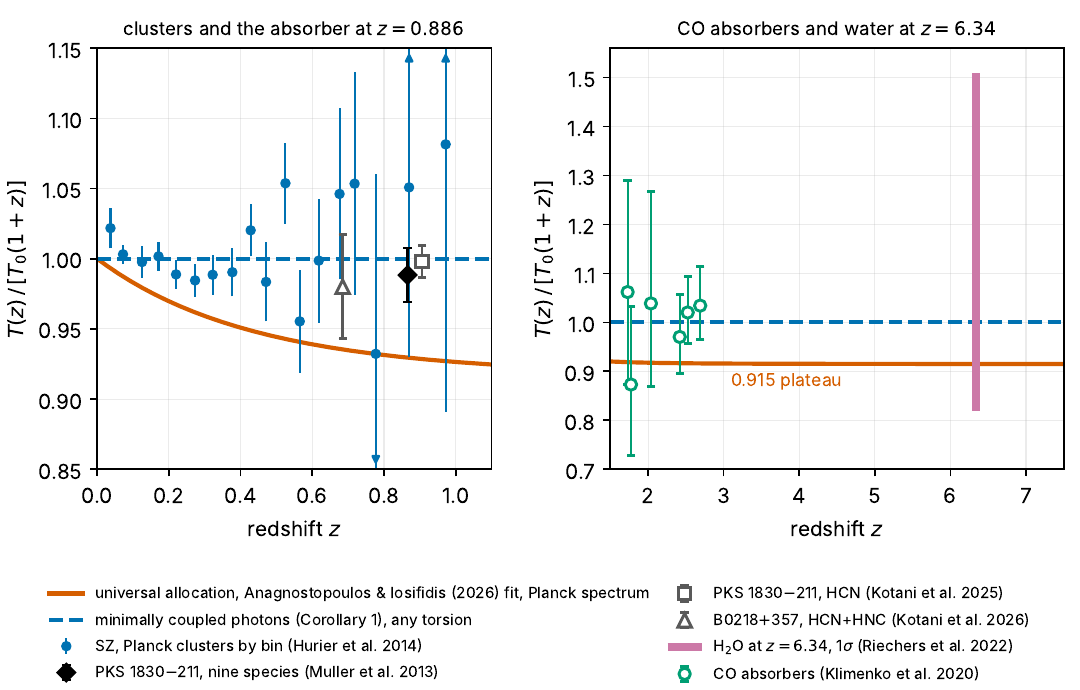}
\caption{CMB temperature relative to $T_{0}(1+z)$, at low redshift (left) and high redshift (right); note the different vertical ranges. Solid line: universal allocation~\eqref{eq:universal} with the central values of the fit of Ref.~\cite{AI2026} ($\Omega_{m0}=0.380$, $n=-1.62$, $h=0.654$); the curve levels off at $0.915$ for $z\gtrsim3$. Dashed line: any EC or metric-affine cosmology with minimally coupled photons (Corollary~\ref{cor:blind}), including the reassignment of Section~\ref{sec:reassignment}. Filled blue circles: SZ measurements in Planck clusters by redshift bin~\cite{Hurier2014}; the error bars shown add statistical and systematic errors in quadrature, whereas the $\chi^{2}$ of the text treats the systematic as fully correlated between bins; arrows mark the bars that extend beyond the frame. Black diamond and open grey square: the two readings of PKS~1830$-$211 (Table~\ref{tab:absorbers}), of the same galaxy, slightly offset in $z$ for visibility and not combined; open grey triangle: B0218$+$357. The HCN and HNC readings assume radiative equilibrium. Open green circles: the six CO systems of Table~\ref{tab:absorbers}. The bar at $z=6.34$ is the $1\sigma$ range from water absorption~\cite{Riechers2022}.}
\label{fig:Tz}
\end{figure}

We integrated the background with the central values of their fit to cosmic chronometers, supernovae and DESI DR2 BAO (their Table~4: $\Omega_{m0}=0.380$, $n=-1.62$, $h=0.654$, with the photon density $\Omega_{r0}=2.47\times10^{-5}h^{-2}$ and no neutrinos, as in their model)~\cite{AI2026}, closing the system with the first of Eqs.~\eqref{eq:AIfried}.\footnote{Ref.~\cite{AI2026} integrates instead a system of three differential equations whose third member is consistent with the Friedmann constraint only when $\ddot{a}=0$; the two closures differ by about 2\% in $H(z)$ at the BAO redshifts and by $0.015\K$ in what follows. \ref{app:closure} gives the details; nothing below depends on the choice.} Figure~\ref{fig:Tz} shows the result. The temperature falls below $T_{0}(1+z)$ from $z=0$ onwards, reaches a 7\% deficit at $z\simeq0.9$ and levels off at $T/[T_{0}(1+z)]\simeq0.915$ for $z\gtrsim3$, where the interaction is negligible. At the redshift of PKS~1830$-$211 the model gives $4.78\K$ ($4.775\K$ at $z=0.886$, with $\pm0.01\K$ from propagating the $1\sigma$ intervals of their Table~4, $\Omega_{m0}=0.380\pm0.008$ and $n=-1.62\pm0.21$, as independent Gaussians).\footnote{The prediction is robust across their parameter space: $T(0.886)$ stays between $4.65$ and $4.85\K$ over the whole box $\Omega_{m0}\in[0.30,0.42]$, $n\in[-2.5,-1.0]$, with a negligible dependence on $h$ through $\Omega_{r0}$ (\ref{app:numbers}); the local slope $\beta(0)=b/2$ depends on $\Omega_{m0}$ alone.}

\subsection{From the balance to the spectrum}
\label{sec:spectrum}

Equation~\eqref{eq:Tz} assumes that the radiation stays a blackbody while it is being fed, and the energy balance alone does not guarantee it. The balance $\dot\rho_{\gamma}+4H\rho_{\gamma}=Q_{\gamma}$ fixes one moment of the photon distribution, its energy. It says nothing about how the new energy is shared among frequencies, and different kinetic operators with the same $Q_{\gamma}$ produce different spectra (\ref{app:kinetic}). The theory does not settle the question; the measured spectrum does, by excluding the alternatives one by one. After recombination there is no process left that can thermalise injected energy, so anything added to the radiation keeps the shape in which it arrived~\cite{ChlubaSunyaev2012} and only redshifts. Here the amount is large: since $T/[T_{0}(1+z)]\rightarrow0.915$ where the interaction is negligible, the model adds to the CMB almost a third of its present energy, at $z\lesssim3$. FIRAS measured the outcome, the spectrum we receive today, between 60 and 600~GHz: a blackbody with residuals below 50 parts per million of the peak, $|y|<1.5\times10^{-5}$ and $|\mu|<9\times10^{-5}$~\cite{FIRAS1996,Fixsen2009}. Strictly, that constrains the redshifted sum of every injection since recombination, not the spectrum at each epoch in between. For a source whose spectral shape is set by local conditions, one mechanism driven by the single amplitude $\sigma(t)$ of the model, the sum is enough. The injection happens while $T/[T_{0}(1+z)]$ moves only between 0.915 and 1, which smears a distortion by less than 9\% in frequency, so any shape of its own (a grey body, a Compton-like redistribution, a spread of temperatures) would leave a distortion two to four orders of magnitude above those limits. The part of the injection that does not go into raising the temperature of a blackbody must then stay below about $3\times10^{-4}$ of the whole, and an injection hidden outside the FIRAS band is excluded by the absolute measurements on either side of it (\ref{app:kinetic}). A source whose shape changed in time, so that distortions injected at one epoch were cancelled by others injected later, is not excluded by FIRAS alone. The model does not propose one, and its energy balance would still fix how much the photons have received by each redshift, only not at which frequencies; Section~\ref{sec:local} says which thermometers survive that freedom. We take the Planck-preserving history~\eqref{eq:Tz} as the model's reading from here on. Either it predicts the temperature of Fig.~\ref{fig:Tz}, or it needs a source tuned to hide its own distortions.

FIRAS fixes the state of the spectrum, not the operator that produces it, and the distinction matters for what follows. Chluba worked out the kinetics of a late, Planck-preserving injection~\cite{Chluba2014}, and \ref{app:kinetic} specialises his analysis to a model in which the amount injected is not free. Two operators reproduce the same history of the blackbody. One rescales every photon momentum by a common factor and drags along anything riding on the spectrum. The other creates photons with the right spectral shape and leaves the photons already present alone; it is the only momentum-local operator that does. They agree on every thermometer that samples the radiation where it sits. They disagree on anything that travels from a source to us: a spectral line, a distortion imprinted by a cluster, the primary anisotropies. And the first is not available to the model. A kinetic operator local in momentum does not know which photons belong to the CMB, so a rescaling would shift the lines by which the redshifts of the fit were measured, breaking the relation $1+z=a_{0}/a$ that Ref.~\cite{AI2026} assumes, and brighten its supernovae by $0.3$~mag at $z\simeq0.9$; expressed in the measured redshift, the absorbers too would then read the standard value. That is a different model, with shifted redshifts and modified distances, which we do not pursue. What is left is creation at the local temperature, which is what the phrase has always meant in decaying-vacuum cosmologies~\cite{Lima1996,LSV2000,Jetzer2011}: the monopole is fed, and everything else propagates as in general relativity. We adopt it, and we say below which part of the test depends on it.

\subsection{Two kinds of thermometer}
\label{sec:local}

Under that reading both kinds of thermometer read $T(z)$, and we keep them apart because they rest on different things. A thermometer that sits in the radiation, like the molecules towards PKS~1830$-$211 or the CO and C\,{\sc i} in quasar absorbers, has its level populations set by the local occupation numbers, and reads the blackbody at its own redshift, independently of how perturbations travel. An SZ thermometer reads a distortion imprinted at the cluster and received today, and needs that distortion to have travelled as in general relativity~\cite{Luzzi2015,Hurier2014,deMartino2015}, which the creation reading guarantees and a rescaling would not (\ref{app:kinetic}).

\begin{table}[t]
\centering\small
\caption{Local thermometers against the universal allocation. $T_{\rm std}=T_{0}(1+z)$; $T_{\rm model}$ from Eq.~\eqref{eq:Tz} with the best fit of Ref.~\cite{AI2026}. The HCN and HNC values assume radiative equilibrium with the CMB (Section~\ref{sec:local}). CO values are the $T_{\rm CMB}$ of Ref.~\cite{Klimenko2020}, Table~2, corrected for collisional excitation.}
\label{tab:absorbers}
\begin{tabular}{@{}llcccc@{}}
\toprule
Absorber & Probe & $z$ & $T$ (K) & $T_{\rm std}$ (K) & $T_{\rm model}$ (K)\\
\midrule
PKS~1830$-$211 & nine species~\cite{Muller2013} & 0.886 & $5.08\pm0.10$ & 5.14 & 4.78\\
PKS~1830$-$211 & HCN~\cite{Kotani2025} & 0.886 & $5.13\pm0.06$ & 5.14 & 4.78\\
B0218$+$357 & HCN, HNC~\cite{Kotani2026} & 0.685 & $4.50\pm0.17$ & 4.59 & 4.30\\
J0857$+$1855 & CO & 1.7294 & $7.9^{+1.7}_{-1.4}$ & 7.44 & 6.83\\
J1047$+$2057 & CO & 1.7738 & $6.6^{+1.2}_{-1.1}$ & 7.56 & 6.94\\
J1705$+$3543 & CO & 2.0377 & $8.6^{+1.9}_{-1.4}$ & 8.28 & 7.60\\
J1439$+$1118 & CO & 2.4184 & $9.04^{+0.8}_{-0.7}$ & 9.32 & 8.54\\
J0000$+$0048 & CO & 2.5244 & $9.8^{+0.7}_{-0.6}$ & 9.61 & 8.80\\
J1237$+$0647 & CO & 2.6896 & $10.4^{+0.8}_{-0.7}$ & 10.06 & 9.21\\
\bottomrule
\end{tabular}
\end{table}

\paragraph*{Absorbers} Table~\ref{tab:absorbers} collects the local thermometers: the two readings of the absorber towards PKS~1830$-$211 at $z=0.886$~\cite{Muller2013,Kotani2025}, the absorber towards B0218$+$357 at $z=0.685$~\cite{Kotani2026}, and the six CO absorption systems at $1.7<z<2.7$ whose rotational excitation Klimenko et al.\ reanalysed with corrections for collisional excitation~\cite{Klimenko2020}, which lower some of the originally published values by up to a kelvin, about ten percent. We take the corrected values, which are the more conservative for us, since the model predicts a colder sky. The absorber towards PKS~1830$-$211 is the sharpest single point: against $5.08\pm0.10\K$~\cite{Muller2013} the model is $3.0\sigma$ low, and against $5.13\pm0.06\K$~\cite{Kotani2025} $5.9\sigma$, the latter conditional on the radiative equilibrium that Kotani et al.\ assume for their HCN lines. The two readings are of the same galaxy, so they are not independent and we do not combine them. The absorber towards B0218$+$357, read with HCN and HNC under the same assumption~\cite{Kotani2026}, gives $4.50\pm0.17\K$ where the standard value is $4.59\K$ and the model $4.30\K$; it sits $0.5\sigma$ below the first and $1.2\sigma$ above the second, and decides nothing by itself. The CO systems are individually less precise, with errors of six to twenty-two percent, but there are six of them, all at redshifts where the model sits on its plateau. Four of the six lie above the standard value, while the model asks all six to lie 8\% below it. With the larger of the two error bars of each point, the six CO systems give $\chi^{2}=1.1$ for the standard scaling and $5.4$ for the universal allocation, neither a bad fit on its own for six points; adding the absorber with the reading of Muller et al.\ gives $1.5$ against $14.7$ for seven points, where $14.7$ has a goodness-of-fit probability of $0.04$; adding B0218$+$357 as well, $1.8$ against $16.1$. The difference, $\Delta\chi^{2}=13$ for the seven points ($15$ with the error bar on the side of each prediction, $14$ with B0218$+$357 included, \ref{app:numbers}), is a likelihood ratio of about $10^{3}$ between two hypotheses with no free parameter; converted by convention into a Gaussian equivalent it is $3.6$ to $3.9\sigma$, and with the reading of Kotani et al.\ in place of Muller's it exceeds $6\sigma$. About $9$ of the $13$ come from PKS~1830$-$211; the six CO systems alone give $\Delta\chi^{2}=4$ to $6$. As a cross-check without a significance attached, the model's effective slope over the CO redshifts is $\beta_{\rm eff}\simeq0.07$--$0.09$, and the absorber-only fit of Klimenko et al., C\,{\sc i} and CO together, gives $\beta=-0.015^{+0.030}_{-0.028}$ (their Table~5)~\cite{Klimenko2020}. At $z=6.34$ water absorption brackets the temperature between 16.4 and 30.2~K~\cite{Riechers2022}; the standard value is $20.0\K$ and the model gives $18.3\K$, both inside.

The two groups of absorbers also differ in what they assume about the spectrum. At the CO redshifts the model has barely started to inject: $T/[T_{0}(1+z)]$ is 0.916 to 0.919 there, against the asymptotic 0.915, so the radiation the molecules see is the blackbody thermalised before recombination, at $0.915\,T_{0}(1+z)$, plus between 0.4 and 1.1\% of the standard energy injected since. That much follows from the energy balance and today's temperature alone. Raising the occupation numbers over the band of the CO rotational transitions, 115 to 346~GHz in the rest frame, to their standard values would take three to four and a half times as much energy. Poured entirely into that band, the energy available would bring it to $0.934$--$0.943\,T_{0}(1+z)$, and the six CO systems would still prefer the standard scaling, by $\Delta\chi^{2}=2.7$ to $3.8$ instead of 4 to 6 (\ref{app:kinetic}). Only a source with narrow features placed at the CO frequencies at every epoch could do better, and such a source would have to know about CO. The absorbers at $z=0.685$ and $0.886$ have no such protection. By $z=0.886$ the model has injected 4.4\% of the standard energy, more than it costs to raise to its standard level any band a factor of two wide below about 270~GHz in the rest frame, where the lowest rotational transitions of HCN and HNC lie. Their reading of $T(z)$, and with it most of the $\Delta\chi^{2}$, rests on the Planck-preserving history.

A word on the systematics of the absorbers, because the largest of them point one way and are of the size of the effect. Every excitation mechanism that is not the CMB (collisions, local radiation, dust, photon trapping) raises the excitation temperature above $T_{\rm CMB}$, in the direction that would hide a deficit. Muller et al.\ give the best defence against contamination. Nine species with different excitation requirements, fitted jointly with the gas density and the kinetic temperature free, agree on a temperature close to $5\K$, which is what coupling to the CMB does and contamination does not~\cite{Muller2013}. The readings of Kotani et al.\ at both absorbers assume radiative equilibrium and neglect collisions~\cite{Kotani2025,Kotani2026}; if collisions contribute, the excitation temperature overestimates $T_{\rm CMB}$, so we quote their $5.9\sigma$ as conditional on that assumption and build the headline on Muller's joint fit. \ref{app:numbers} records what each analysis assumes and why we use the CO systems and not the C\,{\sc i} ones. We take $3\sigma$ as the conservative figure from the absorber, and we say plainly that the absorber test is at the edge of present systematics. A single CO or C\,{\sc i} measurement at the few-percent level, at any $z\gtrsim2$, would settle it.

\paragraph*{Clusters} The SZ thermometers cover $0.04\leq z_{\rm eff}\leq0.97$ with different systematics, and under the creation operator they apply. We compare directly with the 18 redshift bins of Hurier et al.~\cite{Hurier2014}, with $T_{0}=2.7255\K$ and the systematic error treated as the authors prescribe in the notes to their Table~4, ``fully correlated from one redshift bin to the next''. With no free parameter, $\chi^{2}=15.5$ for $T\propto1+z$ and $50.8$ for the universal allocation, for 18 degrees of freedom: the standard scaling is an acceptable fit ($p=0.63$) and the universal allocation is not ($p=5.6\times10^{-5}$, about $4\sigma$ two-sided); treating the systematic as uncorrelated instead gives $14.6$ against $78.0$. The cruder summary says the same thing. A least-squares fit of $T=T_{0}(1+z)^{1-\beta}$, with $T_{0}$ fixed, to the model over the cluster range gives $\beta_{\rm eff}=0.13$, and $0.17$ below $z=0.3$, against $\beta=0.009\pm0.017$ from the Planck clusters of Hurier et al.~\cite{Hurier2014}, $0.012\pm0.016$ from the 103 Planck clusters of Luzzi et al.~\cite{Luzzi2015} and $-0.007\pm0.013$ from the 481 X-ray selected clusters of de Martino et al.~\cite{deMartino2015}. These are formal differences of seven to ten standard deviations (a compilation of all probes gives $\beta=(7.6\pm8.0)\times10^{-3}$~\cite{Avgoustidis2016}), which we quote as indicative only, because a single $\beta$ summarises poorly a model whose slope changes with $z$, and because the cluster samples overlap. One caveat belongs to the clusters alone: their pipelines remove the primary anisotropies with the standard spectral shape, which the creation reading itself modifies (Section~\ref{sec:more}), so the cluster test is not independent of the anisotropy test noted there. The clusters weigh more than the absorbers, but they need one assumption more, about how a perturbation travels, and the absorbers do not. The two agree.

\subsection{What else the model moves}
\label{sec:more}

The temperature is the visible part of something larger, and we mention the rest without using it in the numbers. At $z=1090$ the model has a radiation energy density 30\% below the standard extrapolation (23\% fewer photons) and 16\% less matter. That shifts the redshift of matter--radiation equality upward by about 20\%, and recombination itself, which happens at a fixed temperature, to $1+z\simeq1190$. The Planck anisotropies fix the corresponding quantities to a percent or better~\cite{Planck2018}. A full fit of the universal allocation to the CMB anisotropies would be its most severe test; it would test the whole model, whose best fit already sits far from the Planck $\Lambda$CDM values, and not only the allocation. We have not done it; the thermometer already discriminates at the 3--$4\sigma$ level. There is, besides, a signature in the anisotropies that needs no fit, and it is stronger than any thermometer. Under the creation operator the primary anisotropies, imprinted at recombination and free-streamed since, would today have the spectral shape of the temperature derivative of a Planck function at $0.915\,T_{0}=2.49\K$ rather than at $T_{0}$: relative to 143~GHz, their amplitude would be 5\% higher at 100~GHz and 9\% lower at 217~GHz, where the Planck channels, each calibrated on the dipole, agree at the $10^{-3}$ level between the first and second acoustic peaks~\cite{Planck2018III,Planck2018V}. Chluba derived this effect for small departures and proposed the test~\cite{Chluba2014}; here the departure is 8.5\% (\ref{app:kinetic}). We state it and do not make it the headline, for one reason: it concerns the perturbations of a model whose perturbation theory has not been written, and rests on the transport assumption alone. The clusters share that assumption; the absorbers do not need it, and it is the CO absorbers that isolate the allocation with the fewest hypotheses: the background, today's temperature and the catalogued redshifts. As far as we can see, though, the model has no momentum-local kinetic reading that survives the data.

The baryons raise a related problem. With~\eqref{eq:universal} all dust receives energy, so that the comoving energy of matter today is 19\% larger than at $z=1090$ and 16\% larger than at $z=0.886$. For dark matter that may be a property of the model. For baryons, which are Dirac fermions without vector spin (Section~\ref{sec:noether}), it means creating baryons or changing their mass, and nothing in the model does either.

One earlier work had already used $T(z)$ to bound cosmological torsion. Martins and Vieira constrained the steady-state mode of KTBI, in which torsion is proportional to $H$, through a violation of the distance-duality relation~\cite{MartinsVieira2024} taken from Ref.~\cite{Bolejko2020}, which propagates light with the connection with torsion. With Maxwell's minimal coupling light follows null geodesics of the metric and photon number is conserved, so Etherington's relation holds and $T\propto1+z$ with no correction; read with Proposition~\ref{prop:noether}, their bound measures a non-minimal coupling of photons. In this paper $T(z)$ tests something else, how a model distributes the exchange among its sectors.

\section{Changing the recipient}
\label{sec:reassignment}

What fails is the allocation, not the metric-affine geometry, and the allocation can be changed without touching the rest. Suppose the spin is carried by dark matter, as a phenomenological medium like the hyperdust of AIJS~\cite{AIJS2025,AIJS2026}, and that baryons and photons carry none. Then baryons and photons are conserved (Corollary~\ref{cor:blind}) and the whole of the total exchange (Corollary~\ref{cor:sum}) falls on the dark matter,
\begin{equation}
Q_{c}=\frac{\kappa\sigma}{2}\left(\rho+3p\right),\qquad Q_{b}=Q_{\gamma}=0,
\label{eq:reassignment}
\end{equation}
with the same $\sigma(H)$, the same geometry and the same total exchange~\eqref{eq:AIfried}. The expansion history changes only because the radiation, which no longer receives energy, weighs slightly differently in $\rho+3p$. Integrating both allocations at the fitted parameters, in the same closure, $|E_{\rm reassigned}/E_{\rm universal}-1|\simeq6\times10^{-5}$ at most for $z<3$, which is where the chronometers, supernovae and BAO of the fit live, and the CMB temperature is $T_{0}(1+z)$ again: $5.14\K$ at the absorber. (The split of the dust between baryons and dark matter, $\Omega_{b0}=0.049$, decides who receives the energy but does not enter $E(z)$.) We do not claim more than this comparison shows. It says that the low-redshift data of Ref.~\cite{AI2026} cannot tell the two allocations apart, so the support they give the model they give to both. It does not reproduce the published likelihood, which was computed with a different closure of the background system (\ref{app:closure}), and a refit is theirs to do. The insensitivity below $z=3$ is intrinsic: the radiation is a negligible part of $\rho+3p$ there, so moving its share of the exchange to dark matter changes nothing that the low-redshift data can see. At high redshift the two histories do differ, by close to a percent in $H$ at $z\sim300$ and by 2.6\% at $z=1090$, and the fit of Ref.~\cite{AI2026} is blind to that too, for a different reason: the sound horizon $r_{d}$ is a free parameter there, with a flat prior between 110 and 170~Mpc (their Table~3), rather than a quantity computed from each model's own early history. Computed that way, $r_{d}$ would separate the two allocations.

Corollary~\ref{cor:blind} makes the price explicit. Dark matter must carry vector spin, something minimally coupled Dirac fermions do not have, and its comoving energy grows by 23\% since $z=1090$ (the 19\% of Section~\ref{sec:more} was for all dust; here the baryons are exempt). That figure refers to the energy defined at fixed connection (Section~\ref{sec:noether}) and changes with the bookkeeping. This is the hyperdust of AIJS, the only sector that exchanges energy with the geometry in their setup~\cite{AIJS2026}. There the split is written down; Proposition~\ref{prop:noether} derives it. The reassigned allocation has the standard radiation history, and a fit including the early CMB should use it. Whether such a fit, with $h=0.654$ and $\Omega_{m0}=0.38$, could accommodate the anisotropies is a separate question that we do not address.

\section{Discussion}
\label{sec:discussion}

\paragraph{The rule} In an Einstein--Cartan cosmology the energy the geometry exchanges with matter is not allocated by convention. Each sector receives in proportion to its own spin, and a spinless sector receives nothing. The rule depends only on the invariance of each sector's action under diffeomorphisms and Lorentz transformations, so it holds with any gravitational dynamics. In metric-affine gravity we expect the same to hold with the full hypermomentum, dilation and shear included, in place of the spin; the Noether identities of a matter Lagrangian in that geometry are known~\cite{HMMN1995,ObukhovPuetzfeld2014}, and we have not carried out the reduction here.\footnote{As written in those references the identities have three source terms, torsion against the energy--momentum, non-metricity against the metric energy--momentum and curvature against the hypermomentum; the step that produced~\eqref{eq:diffeo}, splitting the connection into its Levi-Civita part and the distortion, should collect them into derivatives of the distortion against the hypermomentum, as it must if minimally coupled light is to be conserved in a Weyl geometry.} The corollary about photons, the part we set against the data, needs only that Maxwell's action not see the connection, and holds in any of those theories. The thermometers support it in the simplest possible sense: CMB photons behave as if they had no spin, which is what minimal coupling predicts.

\paragraph{The phenomenological literature} Corollary~\ref{cor:nosource} has consequences for a broad family of papers that impose a cosmological torsion by ansatz, proportional to $H$ or to some function of the density, and let the total balance law decide how matter evolves~\cite{KTBI2019,PLJH2019,VJP2025,Liu2025}. Within EC theory that torsion needs a source with spin, and the matter that exchanges energy with it is the matter that produces it. If that is exotic dark matter, there is no contradiction. If the fluid is meant to include baryons or photons, the model is assuming a non-minimal coupling that is not written down, and that would have consequences outside cosmology. The corollary does not reach theories in which torsion propagates or belongs to a different geometry, such as the semi-symmetric metric gravity of Csillag and Harko~\cite{CsillagHarko2024} or the effective treatment of Bolejko et al.~\cite{Bolejko2020}; for those only Corollary~\ref{cor:blind} applies, and it still says that minimally coupled photons are conserved.

\paragraph{Who can carry the spin} The reassignment of Section~\ref{sec:reassignment} asks dark matter for a vector spin, and few candidates can provide one. Not minimally coupled Dirac fermions, whose spin tensor is axial. Not a Weyssenhoff fluid, whose spin $s_{ab}{}^{c}=S_{ab}u^{c}$ with the Frenkel condition $S_{ab}u^{b}=0$ has no vector component in the comoving frame and loses the rest to isotropy, leaving only the contact term of Section~\ref{sec:noether}~\cite{HHK1974,ObukhovKorotky1987}. Nor the axial condensates proposed as microscopic sources of torsion~\cite{CapolupoZeldovich2026}, which source only $Y$. What remains is a phenomenological medium like the hyperfluid of AIJS~\cite{AIJS2025,AIJS2026}, or a non-minimal coupling of the dark sector to the trace of the torsion~\cite{FreidelMinicTakeuchi2005,MZK2013}. A count of magnitudes narrows the options. With the fit of Section~\ref{sec:reassignment}, $|X_{0}|=\kappa\sigma_{c,0}/2\simeq0.4\,H_{0}$, the dark matter must hold today a hypermomentum density $\sigma_{c,0}\simeq0.8\,H_{0}/\kappa\approx7\times10^{21}\,\mathrm{eV}^{3}$, some $10^{36}\hbar$ per cubic centimetre, and a gas of particles of spin $s\hbar$ can supply at most $s\hbar\rho_{c}/m$, so the carrier would have to be lighter than $m\lesssim\tfrac{3}{2}\,s\,\Omega_{c}\,\hbar H_{0}/|X_{0}/H_{0}|\approx\hbar H_{0}\sim10^{-33}\,\mathrm{eV}$; a particle of $1\,\mathrm{GeV}$ falls short by forty-two orders of magnitude.\footnote{A mismatch of the same size was met by Tilquin and Sch\"ucker~\cite{TilquinSchucker2011}, who fitted a vector torsion sourced by the spin of ordinary matter and found the spin per unit energy that the data require to be $10^{42}$ times $\hbar/m_{p}c^{2}$. In EC theory the statement is sharper, because with minimal coupling Dirac matter does not source the vector component at all; a non-minimal coupling multiplies the current of each particle by a dimensionless strength, which would then have to make up the same forty-two orders.} Whatever carries the spin is not a gas of particles diluting with the expansion but a coherent medium whose hypermomentum does not scale with particle number; what the data require of it we take up elsewhere~\cite{WhoFeeds}.

\paragraph{Relation to our previous work} In Ref.~\cite{nogo} we showed that an adiabatic torsion, one that dilutes as $a^{-3}$, has no role in the late universe, and that escaping that conclusion requires the torsion to exchange energy with a source that does not dilute, housed in the matter sector. Proposition~\ref{prop:noether} specifies in which sector: the one that carries the spin. The model of Ref.~\cite{AI2026}, with $\sigma\propto E^{n}$, is not adiabatic and falls outside the hypotheses of that result. Nothing said here argues against torsion as late dark energy, only against an allocation that hands part of the exchange to the photons.

\paragraph{The other end of the history} The same rule has a consequence in the early universe. If torsion is to reheat the universe after inflation, it can only hand energy to sectors with spin: photons have none, and minimally coupled Dirac fermions receive only through the axial channel of~\eqref{eq:Qi}, with $|\zeta_{r}|\leq\hbar n_{r}/2$. Writing $\lambda\equiv|\dot{Y}+HY|/H^{2}$, a fully polarised relativistic fermion bath receives at most $Q_{r}\leq\tfrac{3}{2}\lambda\hbar n_{r}H^{2}$, and its energy density grows only while $T\lesssim0.12\,\lambda\hbar H$ (with $\rho_{r}\simeq3.15\,n_{r}T$ for a Fermi gas without chemical potential; the weaker demand that $\rho_{r}a^{3}$ grow gives $0.48$), where it holds a fraction of order $\lambda^{4}(\hbar H/M_{P})^{2}$ of the critical density. Whether such a bath later comes to dominate depends on how the rest dilutes, which the bound does not control; what it says is that the classical exchange of EC theory with minimally coupled fermions cannot convert the energy that drives the expansion into radiation. A torsion that propagates and is quantised is another matter: it can reheat through decays into the fermionic axial current~\cite{MMN2026}, which is again the channel the rule requires. We develop this, with the bookkeeping of inflation with torsion, in a companion paper~\cite{PaperB}.

\paragraph{Limits} The rule assumes sectors with separate actions, whose dependence on the connection, minimal or not, is written in the action; a non-minimal coupling counts as spin in the sense of~\eqref{eq:currents}. With direct interactions between sectors the energy of each ceases to be defined separately, as in any interacting cosmology, and the rule applies to the whole. The amounts assigned to spinning sectors depend on the energy convention (Section~\ref{sec:noether}); the statement about spinless sectors does not. We have treated classical fields and their isotropic averages, at the level of the background; quantum particle production induced by torsion~\cite{ECportal2021} is a different channel, which we do not consider. The thermometer rests on four further things. The injection must preserve the Planck shape at every epoch, which FIRAS enforces for any source whose spectral shape is set by local conditions, though not for one tuned to cancel its own distortions by today; the CO systems, which see the radiation before most of the injection, lose part of their weight without it but not their sign. The catalogued redshifts must be $1+z=a_{0}/a$, which the fit of Ref.~\cite{AI2026} assumes and which excludes a common rescaling of photon energies. The absorbers must measure the local radiation field, which is how they are calibrated. And, for the clusters only, a perturbation of the spectrum must travel as in general relativity, the creation reading of \ref{app:kinetic}. And in the comparison with data we used only the background history published in Ref.~\cite{AI2026}, taking its radiation to be the CMB, as its own Eq.~(2.51) does. A full fit with the reassignment~\eqref{eq:reassignment}, including the early CMB, remains to be done, and the authors of that reference are better placed than anyone to do it.

\section*{Acknowledgements}

F.~Izaurieta acknowledges support from ANID-FONDECYT Regular 1262414, S.~Lepe from ANID-FONDECYT 1250969, and C.~Quinzacara from ANID-FONDECYT 11231238, Government of Chile. F.~Izaurieta is grateful for the warm hospitality of the Centro de Estudios Cient\'ificos (CECs), Valdivia, where most of this work was done, and for the friendly and stimulating atmosphere of the GeomGravX 2026 conference at the University of Tartu, where discussions on torsion, geometry and cosmology planted the seeds of this work. He also thanks the Netherlands Bach Society for the emotional support of its freely available recordings of Bach, of superb quality, without which this work would have been impossible.

\section*{Data and code availability}
All data used are public and are given in the references cited; the values we use are listed in Table~\ref{tab:absorbers}. Scripts available from the corresponding author.

\section*{Declaration of competing interest}
The authors declare that they have no known competing financial interests or personal relationships that could have appeared to influence the work reported in this paper.

\section*{Declaration of AI-assisted technologies in the writing process}
The authors used Grammarly to correct writing, spelling, grammar, and style errors. They also used the same tool to organize and structure the paragraphs in a clearer and more transparent way. After using this tool, the authors reviewed and edited the text and take full responsibility for the content of the publication.

\appendix

\section{Field equations and identities}
\label{app:forms}

\paragraph{Field equations} Since $\delta_{\omega}R^{ab}=\DD\delta\omega^{ab}$ and $\DD(e^{c}\wedge e^{d})=T^{c}\wedge e^{d}-e^{c}\wedge T^{d}$, the variation of~\eqref{eq:action} with respect to the connection is, after an integration by parts, $\tfrac{1}{2\kappa}\int\epsilon_{abcd}\,\delta\omega^{ab}\wedge T^{c}\wedge e^{d}+\tfrac{1}{2}\int\delta\omega^{ab}\wedge\sigma_{ab}$, from which~\eqref{eq:ec-connection} follows. The variation with respect to the vierbein is $\tfrac{1}{2\kappa}\int\epsilon_{abcd}R^{ab}\wedge e^{c}\wedge\delta e^{d}+\int\delta e^{d}\wedge\tau_{d}$; since a 3-form and a 1-form anticommute, one obtains~\eqref{eq:ec-metric}.

\paragraph{Lorentz identity} Under $\delta e^{a}=\lambda^{a}{}_{b}e^{b}$, $\delta\omega^{ab}=-\DD\lambda^{ab}$, with $\lambda^{ab}=-\lambda^{ba}$ of compact support,
\begin{equation}
0=\delta S_{i}=\int\left(\lambda^{a}{}_{b}\,e^{b}\wedge\tau_{a}-\frac{1}{2}\DD\lambda^{ab}\wedge\sigma_{ab}\right)
=\frac{1}{2}\int\lambda^{ab}\left(-e_{a}\wedge\tau_{b}+e_{b}\wedge\tau_{a}+\DD\sigma_{ab}\right),
\end{equation}
where the term with $\DD\lambda^{ab}$ was integrated by parts and the antisymmetry of $\lambda^{ab}$ was used; since $\lambda^{ab}$ is arbitrary, \eqref{eq:lorentz} follows.

\paragraph{Diffeomorphism identity} The Lie derivative along $\xi$ can be written,\footnote{A historical aside on the contraction operator. Contracting forms with a vector field goes back to Cartan's exterior calculus, which predates the Hodge dual: the latter came with Hodge's work on harmonic integrals in the 1930s~\cite{Hodge1941} and reached most physicists through the book of Flanders~\cite{Flanders1963}. With the dual at hand the contraction can be written with forms alone: in four dimensions and Lorentzian signature, $\II_{\xi}\alpha=-\star(\xi\wedge\star\alpha)$, with $\xi=\xi_{\mu}\dd x^{\mu}$ the 1-form dual to the vector field.} up to a Lorentz transformation with parameter $-\II_{\xi}\omega^{ab}$, as $\delta e^{a}=\DD(\II_{\xi}e^{a})+\II_{\xi}T^{a}$ and $\delta\omega^{ab}=\II_{\xi}R^{ab}$. With $\xi^{a}\equiv\II_{\xi}e^{a}$ of compact support,
\begin{align}
0&=\int\left[\DD\xi^{a}\wedge\tau_{a}+\xi^{b}\left(\II_{b}T^{a}\right)\wedge\tau_{a}+\frac{1}{2}\,\xi^{a}\left(\II_{a}R^{bc}\right)\wedge\sigma_{bc}\right]\nonumber\\
&=\int\xi^{a}\left[-\DD\tau_{a}+\left(\II_{a}T^{b}\right)\wedge\tau_{b}+\frac{1}{2}\left(\II_{a}R^{bc}\right)\wedge\sigma_{bc}\right],
\end{align}
which is~\eqref{eq:hehl}. To pass to the Levi-Civita derivative one uses~\eqref{eq:contortion}: with $\DD\tau_{a}=\dd\tau_{a}-\omega^{b}{}_{a}\wedge\tau_{b}$, $\DD\tau_{a}=\Dr\tau_{a}-K^{b}{}_{a}\wedge\tau_{b}$ and $\II_{a}T^{b}=(\II_{a}K^{b}{}_{c})\,e^{c}-K^{b}{}_{a}$. The two terms in $K^{b}{}_{a}\wedge\tau_{b}$ cancel, leaving $\Dr\tau_{a}=\tfrac{1}{2}(\II_{a}R^{bc})\wedge\sigma_{bc}-(\II_{a}K^{bc})\,e_{b}\wedge\tau_{c}$. Since $K^{bc}$ is antisymmetric only $e_{[b}\wedge\tau_{c]}$ survives, and that is~\eqref{eq:lorentz}.

\paragraph{Fixed-contortion identity} Write the sector action as $S_{i}[e,K,\psi_{i}]\equiv S_{i}[e,\omr(e)+K,\psi_{i}]$. Its variation at fixed $K$ defines $\tau'_{a}$, $\delta S_{i}=\int(\delta e^{a}\wedge\tau'_{a}+\tfrac{1}{2}\delta K^{ab}\wedge\sigma_{ab})$, with $\tau'_{a}=\tau_{a}$ plus the terms that $\delta\omr^{bc}/\delta e^{a}$ brings from $\tfrac{1}{2}\delta\omega^{bc}\wedge\sigma_{bc}$. Under a diffeomorphism combined with the Lorentz transformation of parameter $\II_{\xi}\omr^{ab}$, which removes the term $-(\II_{\xi}\omr^{a}{}_{b})e^{b}$ from the Lie derivative, $\delta e^{a}=\Dr\xi^{a}$ and $\delta K^{ab}=\II_{\xi}\Dr K^{ab}+\Dr(\II_{\xi}K^{ab})$, since $K^{ab}$ is a tensor-valued 1-form and $\omr$ has no torsion. Then
\begin{align}
0&=\int\left[\Dr\xi^{a}\wedge\tau'_{a}+\frac{1}{2}\,\xi^{a}\left(\II_{a}\Dr K^{bc}\right)\wedge\sigma_{bc}+\frac{1}{2}\,\Dr\left(\xi^{a}\II_{a}K^{bc}\right)\wedge\sigma_{bc}\right]\nonumber\\
&=\int\xi^{a}\left[-\Dr\tau'_{a}+\frac{1}{2}\left(\II_{a}\Dr K^{bc}\right)\wedge\sigma_{bc}-\frac{1}{2}\left(\II_{a}K^{bc}\right)\Dr\sigma_{bc}\right],
\end{align}
which is~\eqref{eq:diffeo-contortion}. The same steps with $R^{bc}$ in place of $\Dr K^{bc}$ and $\DD$ in place of $\Dr$ are those of~\eqref{eq:hehl}; the two identities differ by the terms in $\mathring{R}^{bc}$ of $R^{bc}=\mathring{R}^{bc}+\Dr K^{bc}+K^{b}{}_{d}\wedge K^{dc}$ and by terms quadratic in $K$, which are what $\tau'_{a}-\tau_{a}$ absorbs.

\section{Reduction to Friedmann}
\label{app:flrw}

With the vierbein $e^{0}=\dd t$, $e^{j}=a\,\dd x^{j}$ and the torsion~\eqref{eq:torsion-flrw}, the contortion is $K^{0j}=X\,e^{j}$, $K^{jk}=Y\epsilon^{jk}{}_{l}e^{l}$, and since $\omr^{0j}=H\,e^{j}$, $\omr^{jk}=0$, one obtains~\eqref{eq:omega-flrw}; the curvature~\eqref{eq:R-flrw} follows from $R^{ab}=\dd\omega^{ab}+\omega^{a}{}_{c}\wedge\omega^{cb}$. For the component of~\eqref{eq:ec-metric} with $d=0$ only $R^{jk}\wedge e^{l}$ contributes, and with $\epsilon_{jkl0}=-\epsilon_{jkl}$ one gets $-3(Z^{2}-Y^{2})\star e_{0}=\kappa\tau_{0}=-\kappa\rho\star e_{0}$, which is~\eqref{eq:friedmann}; the component with $d=1$ receives from $R^{0j}\wedge e^{k}$ the term with $\dd(aZ)/\dd t$ and from $R^{jk}\wedge e^{0}$ the term with $Z^{2}-Y^{2}$, and gives~\eqref{eq:acceleration}. In~\eqref{eq:ec-connection} the $(0j)$ component contains $T^{k}\wedge e^{l}$ with the $X$ term, and the $(jk)$ component the $Y$ term; with~\eqref{eq:currents-flrw} these give $X=-\tfrac{\kappa}{2}\sum\sigma_{i}$ and $Y=\tfrac{\kappa}{2}\sum\zeta_{i}$. For the identity~\eqref{eq:diffeo}, $\Dr\tau_{0}=\dd\tau_{0}-\omr^{b}{}_{0}\wedge\tau_{b}=-(\dot{\rho}+3H\rho)\,\vol-3Hp\,\vol$, and
\begin{equation}
\frac{1}{2}\left(\II_{0}R^{bc}\right)\wedge\sigma_{bc}=\left(\II_{0}R^{0j}\right)\wedge\sigma_{0j}+\frac{1}{2}\left(\II_{0}R^{jk}\right)\wedge\sigma_{jk}
=\frac{3}{a}\left[\sigma_{i}\,\frac{\dd(aZ)}{\dd t}+\zeta_{i}\,\frac{\dd(aY)}{\dd t}\right]\vol ,
\end{equation}
using $e^{j}\wedge\star e_{j}=3\vol$ and $\epsilon^{jk}{}_{l}\epsilon_{jk}{}^{l}=6$. 

\section{From the energy balance to the photon spectrum}
\label{app:kinetic}

This appendix spells out the kinetic argument behind Section~\ref{sec:spectrum}. Its general part is due to Chluba~\cite{Chluba2014}, who analysed late photon production in the abstract: what a Planck-preserving injection requires, under which conditions the SZ thermometers see it, and what it does to the anisotropies. We specialise his analysis to a model in which the injected fraction is not a small parameter but a fixed number, 30\% of today's energy, and we add the one observation that chooses among the admissible operators: the redshifts the model was fitted to.

\paragraph{What the balance fixes} Let $f(p,t)$ be the isotropic occupation number of the photons, with $p$ the physical momentum. In a Friedmann background,
\begin{equation}
\partial_{t}f-Hp\,\partial_{p}f=C[f],
\label{eq:boltzmann}
\end{equation}
where $C$ collects whatever the exchange with the connection does to the photons. With $\rho_{\gamma}=\pi^{-2}\int p^{3}f\,\dd p$ and $n_{\gamma}=\pi^{-2}\int p^{2}f\,\dd p$ (units $\hbar=c=k_{B}=1$, two polarisations), the moments of~\eqref{eq:boltzmann} are
\begin{equation}
\dot\rho_{\gamma}+4H\rho_{\gamma}=\frac{1}{\pi^{2}}\int p^{3}C\,\dd p\equiv Q_{\gamma},\qquad
\dot n_{\gamma}+3Hn_{\gamma}=\frac{1}{\pi^{2}}\int p^{2}C\,\dd p .
\end{equation}
The balance of Section~\ref{sec:flrw} fixes the third moment of $C$ and nothing else.

\paragraph{What FIRAS adds} A Planck function $f_{\rm Pl}(x)$, $x\equiv p/T(t)$, with $T(t)$ given by~\eqref{eq:Tz}, solves~\eqref{eq:boltzmann} if and only if
\begin{equation}
C[f_{\rm Pl}]=-\left(\frac{\dot T}{T}+H\right)x\,f_{\rm Pl}'(x)=\left(\frac{\dot T}{T}+H\right)x\,f_{\rm Pl}(x)\left[1+f_{\rm Pl}(x)\right]\geq0 ,
\label{eq:CPl}
\end{equation}
a positive source with the spectral shape of the temperature derivative of a Planck function; this is Eq.~(3) of Ref.~\cite{Chluba2014}. Its moments are $Q_{\gamma}=4(\dot T/T+H)\rho_{\gamma}$, which is $\rho_{\gamma}\propto T^{4}$, and $\dot n_{\gamma}+3Hn_{\gamma}=\tfrac{3}{4}(Q_{\gamma}/\rho_{\gamma})n_{\gamma}$, which is $n_{\gamma}\propto T^{3}$: this is the kinetic content of ``adiabatic photon creation'', and, as Chluba stressed, it is a much more special requirement than the conservation of the entropy per photon. Any component of $C$ with a different shape changes the shape of $f$ by an amount set by the energy it carries. Here the energy is $\Delta\rho/\rho\simeq1-0.915^{4}=0.30$ of today's, injected at $z\lesssim3$, when nothing can thermalise it~\cite{ChlubaSunyaev2012}. A Compton-like redistribution of that energy gives $y\simeq\Delta\rho/4\rho\simeq0.075$; a uniform rescaling of the energies of a fixed number of photons, $f\rightarrow\lambda^{-3}f_{\rm Pl}(p/\lambda T)$ with $\lambda-1=\Delta\rho/\rho$, gives a grey body, $\lambda^{-3}$ times a Planck function at temperature $\lambda T$, a departure of order $\Delta\rho/\rho$ at every frequency~\cite{Ellis2013}; a superposition of blackbodies with a temperature spread $\Delta T/T\simeq0.085$ gives a $y$-type distortion of order $\tfrac{1}{2}(\Delta T/T)^{2}\simeq4\times10^{-3}$. Against the FIRAS limits, $|y|<1.5\times10^{-5}$, $|\mu|<9\times10^{-5}$ (95\% confidence) and residuals below $5\times10^{-5}$ of the peak between 60 and 600~GHz~\cite{FIRAS1996,Fixsen2009}, the first two are three to four orders of magnitude too large and the third more than two. For a single injection, the fraction that does not have the shape~\eqref{eq:CPl} must therefore be below $4\times1.5\times10^{-5}/0.30\simeq2\times10^{-4}$ of the whole; the next paragraph says when that extends to a history of injections. An injection placed outside the FIRAS band would not help. Above 600~GHz, 30\% of the energy density of the CMB is about $300\,\mathrm{nW\,m^{-2}\,sr^{-1}}$, several times the whole extragalactic background light. Below 60~GHz lies only 4\% of the energy of a blackbody at $T_{0}$, so the same injection would raise the brightness there by a factor of eight, where ARCADE~2 measures $2.730\pm0.004\K$ between 3 and 90~GHz~\cite{ARCADE2}. We do not derive the operator from the model; we observe that FIRAS leaves~\eqref{eq:CPl} as the only admissible action on the blackbody for a regular source, and read the consequence.

\paragraph{What FIRAS does not fix} FIRAS measures one spectrum, today's. Write the source as the Planck-preserving term~\eqref{eq:CPl} plus a deviation $\delta S(p,t)$ that carries no energy of its own, with $A=B=0$ in the notation of the next paragraph. The solution of~\eqref{eq:boltzmann} today is then
\begin{equation}
f(p,t_{0})=f_{\rm Pl}(p/T_{0})+\int_{t_{\rm rec}}^{t_{0}}\delta S\!\left(p\,\frac{a_{0}}{a(t)},t\right)\dd t ,
\label{eq:fsum}
\end{equation}
and the FIRAS bound applies to the integral, not to $\delta S$ at any one time. If the deviation has a fixed shape in the local variable, $\delta S=q(t)\,h\big(p/T(t)\big)$, the integral is $h$ smeared by the dilations $x\rightarrow x/r$, with $x=p/T_{0}$ and $r(t)=T(t)/[T_{0}(1+z)]$ between 0.915 and 1. A window of 9\% in frequency cannot erase a smooth distortion. In the norm of the intensity, 86\% of a $y$-type, $\mu$-type or grey-body distortion survives it for the model's own injection history, and at least 71\% for any positive weights $q(t)$, so the bound above holds for $h$ itself to within that factor. A source whose whole spectrum is fixed in absolute frequency would instead have to add up, weighted by the model's history, to the Planck difference at $2.7255\K$ to a part in $10^{4}$: a coincidence, not a mechanism.

What FIRAS alone does not exclude is a shape that changes in time, tuned so that distortions injected at one epoch are cancelled by others injected later. Two positive injections with complementary spectra suffice, so no energy needs to be removed. Such a source is not part of the model, but the test should not depend on its absence where it can be avoided, and the energy balance shows where it can. Whatever $\delta S$ is, the energy received by each redshift is fixed. At the redshifts of the six CO systems it is 0.36 to 1.14\% of the standard energy, while raising the occupation numbers over the band of the CO rotational transitions (115.27 to 345.80~GHz in the rest frame) from the primordial blackbody at $0.915\,T_{0}(1+z)$ to the standard one costs 1.63 to 3.57\%, three to four and a half times more. Poured entirely into that band, the energy available would bring it to $0.934$--$0.943\,T_{0}(1+z)$, against $0.916$--$0.919$ in the Planck-preserving reading, and the six CO systems would give $\chi^{2}=3.8$ against $1.1$ for the standard scaling with the larger error bars ($5.1$ against $1.3$ with those towards the prediction). At $z=0.886$ the injected energy is 4.4\% of the standard one, enough to raise to its standard level any band a factor of two wide whose lower edge lies below 136~GHz in the rest frame; at $z=0.685$, 6.5\% and 145~GHz. The absorbers at those redshifts therefore depend on the regularity of the source, and the CO systems, to this extent, do not. Nor do the primary anisotropies, under $A=0$: their perturbation free-streams whatever the monopole does on the way. The clusters do, since the SZ distortion is imprinted on the local spectrum.

\paragraph{Admissible operators} Equation~\eqref{eq:CPl} constrains $C$ on the Planck function only. For a linear operator local in momentum and of first order, $C[f]=A(p,t)\,f+B(p,t)\,\partial_{p}f+S(p,t)$, it is one equation for three functions, and fixes $S$ once $A$ and $B$ are chosen (higher derivatives, of the Kompaneets kind, and the nonlinear Bose factors act on the other photons as well, and change nothing in what follows). What the choice decides is the fate of every photon that is not part of the blackbody. Writing $f=f_{\rm Pl}(p/T(t))+g$, the perturbation obeys $\partial_{t}g-Hp\,\partial_{p}g=A\,g+B\,\partial_{p}g$, whatever $S$ is. The only operator that leaves $g$ to Liouville's equation, transported with $pa$ fixed as in general relativity, is $A=B=0$,
\begin{equation}
C_{\rm cre}[f]=\left(\frac{\dot T}{T}+H\right)x\,f_{\rm Pl}(x)\left[1+f_{\rm Pl}(x)\right],
\label{eq:Ccre}
\end{equation}
a source that does not depend on $f$. Every other choice acts on all photons of momentum $p$, the CMB's or not: $B\neq0$ moves their momenta off $p\propto1/a$, and $A\neq0$ changes their number, which is to say that the exchange acts as a gain or loss medium for whatever light crosses it. The extreme case is the pure drift,
\begin{equation}
C_{\rm resc}[f]=-\left(\frac{\dot T}{T}+H\right)p\,\partial_{p}f,
\label{eq:Cresc}
\end{equation}
which keeps every shape fixed in $x=p/T(t)$, so that all momenta scale as $T(t)$ instead of $1/a$ and the phase-space density is conserved along the modified paths; it creates photons in proportion to those already there, and it would turn any Planck spectrum into the Planck spectrum at the new temperature. Any mixture $\alpha\,C_{\rm resc}+(1-\alpha)\,C_{\rm cre}$ also satisfies~\eqref{eq:CPl}. All of them give the same local Planck function at the same $T(z)$, and any thermometer that samples it through the occupation numbers $f(p=E_{ul})$ at its transition energies reads the same temperature.

\paragraph{Why the rescaling is not available to the model} A kinetic operator local in momentum does not know which photons belong to the CMB. Under $C_{\rm resc}$ the photons of a spectral line emitted at $t_{e}$ arrive with $1+z_{\rm spec}=T(t_{e})/T_{0}=(1+z)\,T(t_{e})/[T_{0}(1+z)]$ instead of $1+z=a_{0}/a$; under the mixture, with $[T(t_{e})/T_{0}(1+z)]^{\alpha}$ in place of the last factor. For the best fit of Ref.~\cite{AI2026} an absorber at $z=0.886$ would be catalogued at $z_{\rm spec}=0.75$, and since its photons gain energy and number on the way, a supernova at that redshift would be brighter by $2.5\log_{10}(0.929^{-4})=0.32$~mag, or $0.32\alpha$~mag for the mixture, with the luminosity distance multiplied by $0.929^{2\alpha}$ at fixed angular-diameter distance. The fit of Ref.~\cite{AI2026} assumes $1+z=a_{0}/a$ for every object and standard luminosity distances, so $C_{\rm resc}$ is not a reading of that model as fitted but a different model. Distance duality is tested at the level of several percent at these redshifts (see Ref.~\cite{Avgoustidis2012} for the link between $T(z)$ and duality under photon non-conservation), which excludes a mixture with $\alpha$ of order one; a small $\alpha$ would again be a different model, with modified distances, which we do not pursue, and for the SZ thermometer it would return $(1-\alpha)$ times the true $\beta$. Two variants deserve a line. An operator confined to a band, acting on millimetre photons and not on optical ones, would make the redshift of an object depend on the band in which its lines are measured; centimetre, millimetre and optical redshifts of the same absorbers agree to parts in $10^{5}$, which is the basis of the radio tests of varying constants~\cite{Uzan2011}. And a source carrying the Bose factor of the actual occupation, $1+f$ in place of $1+f_{\rm Pl}$ in~\eqref{eq:Ccre}, has $A=(\dot T/T+H)\,x f_{\rm Pl}(x)$: a gain medium for the Rayleigh--Jeans part of the CMB band and for nothing else, which amplifies a perturbation by at most $\exp[\,x f_{\rm Pl}(x)\ln(1/0.915)]$, 7\% at 30~GHz and 1\% at 217~GHz, without moving it in frequency. It would tilt the spectral function of a cluster by a few percent; we neglect it. More generally, any operator of the form $C_{\rm cre}+A(p,t)\,[f-f_{\rm Pl}]$ reproduces the same monopole for every $A$ and moves no line, so the redshifts do not constrain $A$. What $A$ is, is a frequency-dependent gain acting on every photon that is not part of the local Planck function: on the distortion imprinted by each cluster, on the primary anisotropies, and on the millimetre continuum of every source between $z\simeq3$ and us. To undo the change of spectral shape of the anisotropies described below, $A$ would have to integrate to the ratio of the two temperature derivatives, a gain of $0.92$ at 30~GHz and $1.10$ at 217~GHz relative to 143~GHz, and the same gain would then be printed on the SZ spectra from which the cluster temperatures are read. Nothing in the balance forbids $A\neq0$; we take $A=0$ as the minimal reading, and the statements about the clusters and the anisotropies are conditional on it. We adopt $C_{\rm cre}$, the reading in which the monopole is fed and everything else propagates as in general relativity.

\paragraph{The two thermometers under each operator} Under $C_{\rm resc}$ the absorbers would read $T(t_{e})=T_{0}(1+z_{\rm spec})$, the standard value in terms of the redshift actually measured, and the whole deviation would sit in the distances. At a cluster, Compton scattering by hot electrons turns the local blackbody into $f_{\rm Pl}(x_{c})+y\,G(x_{c})$, $x_{c}=p/T_{c}$, with $G$ the usual spectral function, and the temperature is inferred from where $G$ sits in frequency when the distortion is received today. Under $C_{\rm resc}$ the perturbation would be carried along with the blackbody, the received occupation would be $y\,G(h\nu_{\rm obs}/k_{B}T_{0})$ whatever $T_{c}$ was, and the fit would return $\beta=0$: the SZ thermometer would measure the temperature of the blackbody in which the distortion sits today, scaled back by $1+z$. This is the blindness Chluba noted for gravitational changes of the temperature, which act on the whole spectrum at once~\cite{Chluba2014}; a kinetic rescaling does the same. Under $C_{\rm cre}$, which is the reading left to the model, the absorbers read $T(z)$ at the catalogued redshift, the perturbation $y\,G$ free-streams, the received occupation is $y\,G(h\nu_{\rm obs}(1+z)/k_{B}T_{c})$, and fitting it with $T_{c}=T_{0}(1+z)^{1-\beta}$ measures the true departure, which is what Refs.~\cite{Luzzi2015,Hurier2014,deMartino2015} assume and do; the relativistic corrections and the kinetic SZ effect, which are also functions of $x_{c}$, free-stream in the same way. That is the division of labour of Section~\ref{sec:local}: both thermometers apply once the rescaling is excluded, and the absorbers then depend on nothing else, save the regularity of the source for those at $z<1$. One more thing ties the clusters to the next paragraph: the SZ pipelines separate the cluster signal from the primary anisotropies using the standard spectral shape of the latter, which under $C_{\rm cre}$ is itself modified.

\paragraph{The primary anisotropies} The same transport has a consequence that needs no thermometer. At recombination the anisotropies are a perturbation $g=\Theta(\hat n)\,x\,f_{\rm Pl}(x)[1+f_{\rm Pl}(x)]$ of the local Planck function, with $x=p/T_{\rm rec}$ and $T_{\rm rec}=0.915\,T_{0}(1+z_{\rm rec})$ in the model. Under $C_{\rm cre}$ it free-streams with $pa$ fixed, and today it has the spectral shape of the temperature derivative of a Planck function at $T_{\rm rec}/(1+z_{\rm rec})=0.915\,T_{0}=2.49\K$, while the monopole, which has been fed, and the kinematic dipole, which is a Doppler shift of the local monopole, have the shapes appropriate to $2.7255\K$. Relative to the standard shape and normalised at 143~GHz, the anisotropy signal would be larger by 9\% at 30~GHz and by 5\% at 100~GHz, and smaller by 9\% at 217~GHz, 27\% at 353~GHz and 46\% at 545~GHz. Chluba derived this effect to first order in the temperature mismatch, as a $y$-type term correlated with the temperature anisotropies (his Eq.~(10a)), bounded the mismatch with the FIRAS dipole temperature on the assumption that the dipole is primordial, which it is not here, and proposed the test with the anisotropies themselves; here the mismatch is 8.5\% and the full shape is needed. Two remarks on robustness. Perturbations of the injection itself, if the spinning medium is inhomogeneous, add photons at each point with the local spectral shape, between those of $0.915\,T_{0}$ and $T_{0}$; that is a late-time, line-of-sight term of the integrated Sachs--Wolfe kind, which cannot restore the standard shape of the acoustic peaks. And the dilution of the anisotropy amplitude by the fed monopole is absorbed in the normalisation of the primordial spectrum, so the test is one of shape alone. The Planck channels, each calibrated on the dipole, show no relative calibration bias between the first and second acoustic peaks at the $10^{-3}$ level~\cite{Planck2018III}, and their cleaned spectra agree to better than one percent~\cite{Planck2018V}, against a predicted 9\% (18\% in power) between 143 and 217~GHz. The creation reading is therefore excluded on its own terms, and so, by the paragraph before last, is the rescaling and any mixture of the two. We know of no operator local in momentum that lets the universal allocation survive. The CO absorbers remain the cleanest statement of the problem, because they use the energy balance, today's temperature and the catalogued redshifts, and still point the same way under the most favourable redistribution of the injected energy.

\section{On the closure of the background system of Ref.~\texorpdfstring{\cite{AI2026}}{[19]}}
\label{app:closure}

The quotations in the main text are from the abstract and from Sec.~2.2 (p.~7) of the arXiv version of Ref.~\cite{AI2026} (v1). The background equations of the pure-spin hyperfluid are their system~(2.24),
\begin{align}
3H^{2}&=\kappa\rho+3H\kappa\sigma-\frac{3\kappa^{2}\sigma^{2}}{4},\qquad
2\dot{H}+3H^{2}=-\kappa p+2H\kappa\sigma+\dot{\sigma}\kappa-\frac{\kappa^{2}\sigma^{2}}{4},\nonumber\\
\dot{\rho}+3H(\rho+p)&=\frac{\kappa\sigma}{2}(\rho+3p),
\label{eq:AI224}
\end{align}
which are~\eqref{eq:friedmann}, \eqref{eq:acceleration} and Corollary~\ref{cor:sum} with $Y=0$ and $X=-\kappa\sigma/2$. The dark energy is identified with the spin terms, their Eqs.~(2.28)--(2.29),
\begin{equation}
\rho_{\rm DE}=3H\sigma-3\kappa\sigma^{2}/4,\qquad P_{\rm DE}=-2H\sigma-\dot{\sigma}+\kappa\sigma^{2}/4,
\label{eq:AI228}
\end{equation}
and the universal coupling is written for the three components, their Eqs.~(2.33)--(2.35),
\begin{equation}
\dot{\rho}_{m}+3H\rho_{m}=\kappa\frac{\sigma}{2}\rho_{m},\qquad
\dot{\rho}_{r}+4H\rho_{r}=\kappa\sigma\rho_{r},\qquad
\dot{\rho}_{\rm DE}+3H(1+w_{\rm DE})\rho_{\rm DE}=\kappa\frac{\sigma}{2}(1+3w_{\rm DE})\rho_{\rm DE}.
\label{eq:AI233}
\end{equation}
With $\kappa\sigma/2H_{0}=\beta E^{n}$, their Eq.~(2.58), where their $\beta$ is our $b$, the first of Eqs.~\eqref{eq:AI224} divided by $3H_{0}^{2}$ gives their Eq.~(2.59),
\begin{equation}
\Omega_{\rm DE}=\frac{\kappa\rho_{\rm DE}}{3H_{0}^{2}}=2\beta E^{n+1}-\beta^{2}E^{2n},
\label{eq:AI259}
\end{equation}
that is, $E^{2}=\Omega_{m}+\Omega_{r}+2bE^{n+1}-b^{2}E^{2n}$, or $(E-bE^{n})^{2}=\Omega_{m}+\Omega_{r}$. We close the background with this relation, on the branch $Z=H-\kappa\sigma/2>0$ that contains the fit ($1-b=\sqrt{\Omega_{m0}+\Omega_{r0}}>0$), which fixes $E(z)$ algebraically once $\Omega_{m}(z)$ and $\Omega_{r}(z)$ are known: on this branch $bE^{n-1}<1$, so $\dd(E-bE^{n})/\dd E=1-nbE^{n-1}>1-n>0$ for every $n<1$, which covers their whole prior, and the root is unique at every redshift. The same relation fixes the future of the model under either allocation: as $\rho\rightarrow0$ it gives $H\rightarrow\kappa\sigma/2$, that is, $Z\rightarrow0$, the torsion-sustained de Sitter state of KTBI~\cite{KTBI2019}, which the companion paper~\cite{PaperB} meets as the end state of a drained early dark energy; with the polytropic closure the asymptotic rate is $E_{\infty}=b^{1/(1-n)}$, about $0.69$ for the best fit.

Ref.~\cite{AI2026} integrates instead the system of their Eqs.~(2.53)--(2.55),
\begin{align}
\frac{\dd\Omega_{m}}{\dd z}&=\frac{\Omega_{m}}{1+z}\left(3-bE^{n-1}\right),\qquad
\frac{\dd\Omega_{r}}{\dd z}=\frac{\Omega_{r}}{1+z}\left(4-2bE^{n-1}\right),\nonumber\\
\frac{\dd\Omega_{\rm DE}}{\dd z}&=\frac{1}{1-\Omega_{\rm DE}C_{1}}\left[\frac{\Omega_{\rm DE}C_{0}}{1+z}+\Omega_{\rm DE}C_{1}\left(\frac{\dd\Omega_{m}}{\dd z}+\frac{\dd\Omega_{r}}{\dd z}\right)\right],
\label{eq:AI253}
\end{align}
with $C_{0}=2/(2-bE^{n-1})$ and $C_{1}=n(1-bE^{n-1})/[(2-bE^{n-1})\Omega_{\rm tot}]$, the third of which is the last of Eqs.~\eqref{eq:AI233} rewritten in redshift. That last equation is not implied by~\eqref{eq:AI224}, and the check takes three lines. (The third of Eqs.~\eqref{eq:AI224} is: it follows from the first two by the Bianchi identity.) With $\rho_{\rm DE}$ and $P_{\rm DE}$ defined by~\eqref{eq:AI228}, the first two of Eqs.~\eqref{eq:AI224} read $3H^{2}=\kappa(\rho+\rho_{\rm DE})$ and $2\dot{H}+3H^{2}=-\kappa(p+P_{\rm DE})$, so the Bianchi identity of the metric sector makes $\rho+\rho_{\rm DE}$ conserved and the third of Eqs.~\eqref{eq:AI224} gives $\dot\rho_{\rm DE}+3H(\rho_{\rm DE}+P_{\rm DE})=-\tfrac{\kappa\sigma}{2}(\rho+3p)$. The last of Eqs.~\eqref{eq:AI233} has $+\tfrac{\kappa\sigma}{2}(\rho_{\rm DE}+3P_{\rm DE})$ on the right instead. The two agree if and only if $\rho+3p+\rho_{\rm DE}+3P_{\rm DE}=0$, that is, $\ddot a=0$: the system~\eqref{eq:AI253} is consistent with the constraint~\eqref{eq:AI259} only in a universe that neither accelerates nor decelerates. The same conclusion follows if the $\rho$ of the third of Eqs.~\eqref{eq:AI224} is read as including $\rho_{\rm DE}$, as the sentence after their Eq.~(2.27) might suggest: that equation is then the sum of~\eqref{eq:AI233} and contradicts the Bianchi identity by the same term, while the first of Eqs.~\eqref{eq:AI224} cannot include $\rho_{\rm DE}$ without counting the spin terms twice. Integrated as written in an accelerating one, it drifts away from the relation that its own Friedmann equation imposes. Table~\ref{tab:drift} gives the size of the drift; it reaches $0.08$ in $\Omega_{\rm DE}$ near $z\simeq0.65$ and about 2\% in $H$ at the BAO redshifts. Eq.~\eqref{eq:AI259} itself provides the consistent closure, and it is the one we use.

\begin{table}[t]
\centering\small
\caption{Difference between the system~\eqref{eq:AI253}, integrated as written, and the algebraic closure~\eqref{eq:AI259}, at the best-fit parameters of Ref.~\cite{AI2026}.}
\label{tab:drift}
\begin{tabular}{@{}lccccc@{}}
\toprule
$z$ & 0.30 & 0.51 & 0.71 & 0.93 & 1.32\\
\midrule
$H_{\rm ODE}/H_{\rm alg}-1$ & $+2.2\%$ & $+2.3\%$ & $+1.9\%$ & $+1.5\%$ & $+0.8\%$\\
\bottomrule
\end{tabular}
\end{table}

For the temperature this is immaterial: with their system in place of the constraint the absorber would read $4.79\K$ and the tensions of Section~\ref{sec:local} would be $2.9\sigma$ and $5.7\sigma$. The parameters of their Table~4 were obtained with the drifting system; we expect a refit with the constraint to move them by less than the parameter box of \ref{app:numbers}, but we have not done it. This is also why Section~\ref{sec:reassignment} compares the two allocations within one closure rather than against the published likelihood: the comparison shows that the allocation is invisible to the low-redshift data, not that the published likelihood is reproduced.

\section{Numerical details and reproducibility}
\label{app:numbers}

\paragraph{Background} For the model of Ref.~\cite{AI2026} we solved the first of Eqs.~\eqref{eq:AIfried} in the form $E-bE^{n}=\sqrt{\Omega_{m}(z)+\Omega_{r}(z)}$, choosing the root that gives $E=1$ today, and integrated it with~\eqref{eq:universal} or with~\eqref{eq:reassignment} in the variable $\ln(1+z)$ up to $z=10^{4}$, with relative tolerances of $10^{-11}$. The temperature follows from $\rho_{r}\propto T^{4}$. At $z=1090$ the universal allocation gives $\rho_{r}/[\rho_{r0}(1+z)^{4}]=0.701$ and $\rho_{m}/[\rho_{m0}(1+z)^{3}]=0.837$, that is, 30\% less radiation and 16\% less matter than the standard extrapolation of today's densities; read the other way, the comoving energies have grown since then by 43\% and 19\%. Over the parameter box quoted in Section~\ref{sec:thermometer}, $\Omega_{m0}\in[0.30,0.42]$ and $n\in[-2.5,-1.0]$, the root of the constraint is unique at every redshift and $T(0.886)$ varies monotonically in both parameters, from $4.65\K$ to $4.85\K$.

\paragraph{Absorbers} The values of Table~\ref{tab:absorbers} are taken as published; for the CO systems we use the $T_{\rm CMB}$ column of Table~2 of Ref.~\cite{Klimenko2020}, which includes the correction for collisional excitation, rather than the excitation temperatures; the corrections reach one kelvin at $z\simeq2$, comparable to the 8\% at stake. Klimenko et al.\ also give $T_{\rm CMB}$ from the fine-structure levels of C\,{\sc i} in nine systems (their Table~4): those levels are pumped by the ultraviolet field and by collisions with hydrogen as well as by the CMB, the correction is larger and more model-dependent than for CO, five of the nine values are upper limits and the other four have lower errors of 30 to 60 percent; their C\,{\sc i}-only fit gives $\beta=-0.077^{+0.130}_{-0.075}$ against $-0.007^{+0.030}_{-0.031}$ for CO alone (their Table~5). We therefore use the CO systems only; the absorber-only $\beta$ quoted in Section~\ref{sec:local} combines both. Muller et al.\ fit their nine species jointly with a non-LTE excitation code, with the gas density and the kinetic temperature as free parameters alongside $T_{\rm CMB}$, assuming a homogeneous medium that covers the source;\footnote{Nine species enter the joint fit; Kotani et al.\ count ten, with ortho and para c-C$_{3}$H$_{2}$ separately.} the densities they derive are of a few hundred to a few thousand molecules per cubic centimetre, at which the species with low critical densities are barely affected by collisions, while SiO and HNCO, which need denser gas, come out warmer and are kept in the fit with their collisional excitation modelled~\cite{Muller2013}. The value $5.08\pm0.10\K$ is the statistical interval of that joint fit, not an average of rotation temperatures, which individually range from about $4$ to $10\K$; it does not include the covering factor or clumpiness, which the authors fix. Kotani et al.\ select HCN transitions for which they argue that collisional excitation is negligible, and assume radiative equilibrium with the CMB~\cite{Kotani2025}; their $5.13\pm0.06\K$ is a measurement conditional on that assumption, and if collisions do contribute it overestimates $T_{\rm CMB}$, as excitation temperatures generally do~\cite{Chluba2014}. The same assumption underlies their value for B0218$+$357, where HCO$^{+}$ was excluded because of collisional excitation~\cite{Kotani2026}; with the larger error bar, adding that absorber to the seven points gives $\chi^{2}=1.8$ for the standard scaling and $16.1$ for the universal allocation. Not every systematic is one-sided: a covering factor below unity, line saturation, the placement of the continuum and the variability of the background source can bias the inferred optical depths either way. For asymmetric errors we quote two conventions: the error bar on the side of the prediction being tested, which is the likelihood choice when the two sides are treated as half-Gaussians, and the larger of the two, which is the conservative one. Table~\ref{tab:chi2} gives the $\chi^{2}$ values. The significances quoted in Section~\ref{sec:local} convert $\Delta\chi^{2}$ between the two hypotheses, which have no free parameter, into a two-sided Gaussian equivalent; the likelihood ratio is $e^{\Delta\chi^{2}/2}$. The effective slope of the model at the CO redshifts, $\beta_{\rm eff}=-\ln\{T/[T_{0}(1+z)]\}/\ln(1+z)$, ranges from $0.067$ to $0.085$.

\begin{table}[t]
\centering\small
\caption{$\chi^{2}$ of the two parameter-free hypotheses against the local thermometers, for the two conventions on asymmetric errors. In the row ``redistributed'' the energy the model has injected by each CO redshift is moved entirely into the band of the CO transitions, the most favourable case for the universal allocation (\ref{app:kinetic}).}
\label{tab:chi2}
\setlength{\tabcolsep}{4pt}
\begin{tabular}{@{}lcccc@{}}
\toprule
 & \multicolumn{2}{c}{error bar towards the prediction} & \multicolumn{2}{c}{larger error bar}\\
Data set (points) & standard & universal & standard & universal\\
\midrule
six CO systems (6) & 1.3 & 7.3 & 1.1 & 5.4\\
\quad CO, redistributed (\ref{app:kinetic}) & 1.3 & 5.1 & 1.1 & 3.8\\
CO $+$ PKS~1830$-$211, Muller et al.\ (7) & 1.6 & 16.6 & 1.5 & 14.7\\
\quad the same $+$ B0218$+$357 (8) & 1.9 & 18.1 & 1.8 & 16.1\\
CO $+$ PKS~1830$-$211, Kotani et al.\ (7) & 1.3 & 42.4 & 1.2 & 40.4\\
\bottomrule
\end{tabular}
\end{table}

\paragraph{Clusters} We used the 18 bins of Table~4 of Ref.~\cite{Hurier2014}, with covariance $C_{jk}=\sigma^{\rm stat}_{j}{}^{2}\delta_{jk}+\sigma^{\rm sys}_{j}\sigma^{\rm sys}_{k}$ and $T_{0}=2.7255\K$. The published table lists the bin edges and the temperatures with their statistical and systematic errors; we place each bin at the effective redshift of its clusters as listed in the arXiv version of the paper (Table~3 of arXiv:1311.4694; the last bin at $z=0.972$), which reproduces the authors' own goodness of fit for the adiabatic law ($\chi^{2}/{\rm dof}=0.86$ against their $0.89$). Placing the bins at their centres instead gives $\chi^{2}=21.1$ and $51.6$. With the systematic added in quadrature, as if uncorrelated, the values are $14.6$ and $78.0$, and without it $15.5$ and $84.9$; full correlation is therefore the conservative choice. As a control, fitting $\beta$ to the same bins and covariance gives $\beta=0.014\pm0.022$; with the statistical errors alone, which is how Hurier et al.\ fit $\beta$, since a systematic fully correlated between bins does not affect it (their Sec.~7.2), we obtain $0.0095\pm0.017$, their published $0.009\pm0.017$. The value $\beta_{\rm eff}=0.13$ for the model is a least-squares fit of $\ln T$ with $T_{0}$ fixed and uniform weighting in $z$ over $0.01\leq z\leq0.97$; a weighting concentrated at $z\sim0.2$, like that of the Planck clusters, gives about $0.15$, and the range $z\leq0.3$ gives $0.17$.

\bibliographystyle{elsarticle-num}
\bibliography{TorsionSpin_refs}

@article{Muller2013,
  author = "Muller, S. and Beelen, A. and Black, J. H. and Curran, S. J. and Horellou, C. and Aalto, S. and Combes, F. and Gu{\'e}lin, M. and Henkel, C.",
  title = "{A precise and accurate determination of the cosmic microwave background temperature at $z=0.89$}",
  journal = "Astron. Astrophys.",
  volume = "551",
  pages = "A109",
  year = "2013",
  doi = "10.1051/0004-6361/201220613",
  eprint = "1212.5456",
  archivePrefix = "arXiv",
  primaryClass = "astro-ph.CO"
}

@article{Kotani2025,
  author = "Kotani, T. and Oka, T. and Enokiya, R. and Yanagihara, K. and Kaneko, M. and Ariyama, R.",
  title = "{A new precise measurement of the cosmic microwave background radiation temperature at $z=0.89$ toward PKS1830$-$211}",
  journal = "Astrophys. J.",
  volume = "993",
  pages = "134",
  year = "2025",
  doi = "10.3847/1538-4357/ae0a1e",
  eprint = "2509.20760",
  archivePrefix = "arXiv"
}

@article{Fixsen2009,
  author = "Fixsen, D. J.",
  title = "{The temperature of the cosmic microwave background}",
  journal = "Astrophys. J.",
  volume = "707",
  pages = "916",
  year = "2009",
  doi = "10.1088/0004-637X/707/2/916",
  eprint = "0911.1955",
  archivePrefix = "arXiv",
  primaryClass = "astro-ph.CO"
}

@article{Klimenko2020,
  author = "Klimenko, V. V. and Ivanchik, A. V. and Petitjean, P. and Noterdaeme, P. and Srianand, R.",
  title = "{Estimation of the cosmic microwave background temperature from atomic C\,{\sc i} and molecular CO lines in the interstellar medium of early galaxies}",
  journal = "Astron. Lett.",
  volume = "46",
  pages = "715",
  year = "2020",
  doi = "10.1134/S1063773720110031",
  eprint = "2106.00119",
  archivePrefix = "arXiv",
  primaryClass = "astro-ph.CO"
}

@article{Noterdaeme2011,
  author = "Noterdaeme, P. and Petitjean, P. and Srianand, R. and Ledoux, C. and L{\'o}pez, S.",
  title = "{The evolution of the cosmic microwave background temperature: Measurements of $T_{\rm CMB}$ at high redshift from carbon monoxide excitation}",
  journal = "Astron. Astrophys.",
  volume = "526",
  pages = "L7",
  year = "2011",
  doi = "10.1051/0004-6361/201016140",
  eprint = "1012.3164",
  archivePrefix = "arXiv",
  primaryClass = "astro-ph.CO"
}

@article{Riechers2022,
  author = "Riechers, D. A. and Weiss, A. and Walter, F. and Carilli, C. L. and Cox, P. and Decarli, R. and Neri, R.",
  title = "{Microwave background temperature at a redshift of 6.34 from H$_2$O absorption}",
  journal = "Nature",
  volume = "602",
  pages = "58",
  year = "2022",
  doi = "10.1038/s41586-021-04294-5",
  eprint = "2202.00693",
  archivePrefix = "arXiv",
  primaryClass = "astro-ph.GA"
}

@article{Hurier2014,
  author = "Hurier, G. and Aghanim, N. and Douspis, M. and Pointecouteau, E.",
  title = "{Measurement of the $T_{\rm CMB}$ evolution from the Sunyaev--Zel'dovich effect}",
  journal = "Astron. Astrophys.",
  volume = "561",
  pages = "A143",
  year = "2014",
  doi = "10.1051/0004-6361/201322632",
  eprint = "1311.4694",
  archivePrefix = "arXiv",
  primaryClass = "astro-ph.CO"
}

@article{Luzzi2015,
  author = "Luzzi, G. and G{\'e}nova-Santos, R. T. and Martins, C. J. A. P. and De Petris, M. and Lamagna, L.",
  title = "{Constraining the evolution of the CMB temperature with SZ measurements from Planck data}",
  journal = "JCAP",
  volume = "09",
  pages = "011",
  year = "2015",
  doi = "10.1088/1475-7516/2015/09/011",
  eprint = "1502.07858",
  archivePrefix = "arXiv",
  primaryClass = "astro-ph.CO"
}

@article{deMartino2015,
  author = "de Martino, I. and G{\'e}nova-Santos, R. and Atrio-Barandela, F. and Ebeling, H. and Kashlinsky, A. and Kocevski, D. and Martins, C. J. A. P.",
  title = "{Constraining the redshift evolution of the cosmic microwave background blackbody temperature with Planck data}",
  journal = "Astrophys. J.",
  volume = "808",
  pages = "128",
  year = "2015",
  doi = "10.1088/0004-637X/808/2/128",
  eprint = "1502.06707",
  archivePrefix = "arXiv",
  primaryClass = "astro-ph.CO"
}

@article{Hehl1976,
  author = "Hehl, F. W. and von der Heyde, P. and Kerlick, G. D. and Nester, J. M.",
  title = "{General relativity with spin and torsion: Foundations and prospects}",
  journal = "Rev. Mod. Phys.",
  volume = "48",
  pages = "393",
  year = "1976",
  doi = "10.1103/RevModPhys.48.393"
}

@article{HMMN1995,
  author = "Hehl, F. W. and McCrea, J. D. and Mielke, E. W. and Ne'eman, Y.",
  title = "{Metric-affine gauge theory of gravity: field equations, Noether identities, world spinors, and breaking of dilation invariance}",
  journal = "Phys. Rept.",
  volume = "258",
  pages = "1",
  year = "1995",
  doi = "10.1016/0370-1573(94)00111-F",
  eprint = "gr-qc/9402012",
  archivePrefix = "arXiv",
  primaryClass = "gr-qc"
}

@article{ObukhovPuetzfeld2014,
  author = "Obukhov, {\relax Yu}. N. and Puetzfeld, D.",
  title = "{Conservation laws in gravity: A unified framework}",
  journal = "Phys. Rev. D",
  volume = "90",
  pages = "024004",
  year = "2014",
  doi = "10.1103/PhysRevD.90.024004",
  eprint = "1405.4003",
  archivePrefix = "arXiv",
  primaryClass = "gr-qc"
}

@article{ObukhovKorotky1987,
  author = "Obukhov, {\relax Yu}. N. and Korotky, V. A.",
  title = "{The Weyssenhoff fluid in Einstein--Cartan theory}",
  journal = "Class. Quantum Grav.",
  volume = "4",
  pages = "1633",
  year = "1987",
  doi = "10.1088/0264-9381/4/6/021"
}

@article{Gasperini1986,
  author = "Gasperini, M.",
  title = "{Spin-dominated inflation in the Einstein--Cartan theory}",
  journal = "Phys. Rev. Lett.",
  volume = "56",
  pages = "2873",
  year = "1986",
  doi = "10.1103/PhysRevLett.56.2873"
}

@article{KTBI2019,
  author = "Kranas, D. and Tsagas, C. G. and Barrow, J. D. and Iosifidis, D.",
  title = "{Friedmann-like universes with torsion}",
  journal = "Eur. Phys. J. C",
  volume = "79",
  pages = "341",
  year = "2019",
  doi = "10.1140/epjc/s10052-019-6822-4",
  eprint = "1809.10064",
  archivePrefix = "arXiv"
}

@article{Iosifidis2020,
  author = "Iosifidis, D.",
  title = "{Cosmological hyperfluids, torsion and non-metricity}",
  journal = "Eur. Phys. J. C",
  volume = "80",
  pages = "1042",
  year = "2020",
  doi = "10.1140/epjc/s10052-020-08634-z",
  eprint = "2003.07384",
  archivePrefix = "arXiv",
  primaryClass = "gr-qc"
}

@article{AIJS2025,
  author = "Andrei, I. and Iosifidis, D. and J{\"a}rv, L. and Saal, M.",
  title = "{Friedmann cosmology with hyperfluids}",
  journal = "Phys. Rev. D",
  volume = "111",
  pages = "064063",
  year = "2025",
  doi = "10.1103/PhysRevD.111.064063",
  eprint = "2411.19127",
  archivePrefix = "arXiv",
  primaryClass = "gr-qc"
}

@article{DESIDR2,
  author = "{DESI Collaboration} and Abdul-Karim, M. and others",
  title = "{DESI DR2 results. II. Measurements of baryon acoustic oscillations and cosmological constraints}",
  journal = "Phys. Rev. D",
  volume = "112",
  pages = "083515",
  year = "2025",
  doi = "10.1103/tr6y-kpc6",
  eprint = "2503.14738",
  archivePrefix = "arXiv",
  primaryClass = "astro-ph.CO"
}

@misc{AI2026,
  author = "Anagnostopoulos, F. K. and Iosifidis, D.",
  title = "{Interacting fluid cosmologies from the metric-affine framework}",
  year = "2026",
  eprint = "2609.07978",
  archivePrefix = "arXiv"
}

@article{AIJS2026,
  author = "Andrei, I. and Iosifidis, D. and J{\"a}rv, L. and Saal, M.",
  title = "{Friedmann cosmology with fluids and hyperfluids}",
  journal = "J. Phys.: Conf. Ser.",
  volume = "3177",
  pages = "012033",
  year = "2026",
  doi = "10.1088/1742-6596/3177/1/012033",
  eprint = "2604.03343",
  archivePrefix = "arXiv",
  primaryClass = "gr-qc"
}

@article{Wang2016,
  author = "Wang, B. and Abdalla, E. and Atrio-Barandela, F. and Pav{\'o}n, D.",
  title = "{Dark matter and dark energy interactions: theoretical challenges, cosmological implications and observational signatures}",
  journal = "Rept. Prog. Phys.",
  volume = "79",
  pages = "096901",
  year = "2016",
  doi = "10.1088/0034-4885/79/9/096901",
  eprint = "1603.08299",
  archivePrefix = "arXiv",
  primaryClass = "astro-ph.CO"
}

@article{Amendola2000,
  author = "Amendola, L.",
  title = "{Coupled quintessence}",
  journal = "Phys. Rev. D",
  volume = "62",
  pages = "043511",
  year = "2000",
  doi = "10.1103/PhysRevD.62.043511",
  eprint = "astro-ph/9908023",
  archivePrefix = "arXiv",
  primaryClass = "astro-ph"
}

@article{Pourtsidou2013,
  author = "Pourtsidou, A. and Skordis, C. and Copeland, E. J.",
  title = "{Models of dark matter coupled to dark energy}",
  journal = "Phys. Rev. D",
  volume = "88",
  pages = "083505",
  year = "2013",
  doi = "10.1103/PhysRevD.88.083505",
  eprint = "1307.0458",
  archivePrefix = "arXiv",
  primaryClass = "astro-ph.CO"
}

@article{Bohmer2015,
  author = "B{\"o}hmer, C. G. and Tamanini, N. and Wright, M.",
  title = "{Interacting quintessence from a variational approach. I. Algebraic couplings}",
  journal = "Phys. Rev. D",
  volume = "91",
  pages = "123002",
  year = "2015",
  doi = "10.1103/PhysRevD.91.123002",
  eprint = "1501.06540",
  archivePrefix = "arXiv",
  primaryClass = "gr-qc"
}

@incollection{HehlObukhov2001,
  author = "Hehl, F. W. and Obukhov, {\relax Yu}. N.",
  title = "{How does the electromagnetic field couple to gravity, in particular to metric, nonmetricity, torsion, and curvature?}",
  booktitle = "{Gyros, Clocks, Interferometers\ldots: Testing Relativistic Gravity in Space}",
  editor = "L{\"a}mmerzahl, C. and Everitt, C. W. F. and Hehl, F. W.",
  series = "Lect. Notes Phys.",
  volume = "562",
  publisher = "Springer",
  address = "Berlin",
  pages = "479--504",
  year = "2001",
  doi = "10.1007/3-540-40988-2_25",
  eprint = "gr-qc/0001010",
  archivePrefix = "arXiv",
  primaryClass = "gr-qc"
}

@article{PuetzfeldObukhov2008,
  author = "Puetzfeld, D. and Obukhov, {\relax Yu}. N.",
  title = "{Probing non-Riemannian spacetime geometry}",
  journal = "Phys. Lett. A",
  volume = "372",
  pages = "6711",
  year = "2008",
  doi = "10.1016/j.physleta.2008.09.041",
  eprint = "0708.1926",
  archivePrefix = "arXiv",
  primaryClass = "gr-qc"
}

@article{BHL2008,
  author = "Brechet, S. D. and Hobson, M. P. and Lasenby, A. N.",
  title = "{Classical big-bounce cosmology: dynamical analysis of a homogeneous and irrotational Weyssenhoff fluid}",
  journal = "Class. Quantum Grav.",
  volume = "25",
  pages = "245016",
  year = "2008",
  doi = "10.1088/0264-9381/25/24/245016",
  eprint = "0807.2523",
  archivePrefix = "arXiv"
}

@article{Poplawski2010,
  author = "Pop{\l}awski, N. J.",
  title = "{Cosmology with torsion: An alternative to cosmic inflation}",
  journal = "Phys. Lett. B",
  volume = "694",
  pages = "181",
  year = "2010",
  doi = "10.1016/j.physletb.2010.09.056",
  eprint = "1007.0587",
  archivePrefix = "arXiv",
  primaryClass = "astro-ph.CO"
}

@article{ILV2020,
  author = "Izaurieta, F. and Lepe, S. and Valdivia, O.",
  title = "{The spin tensor of dark matter and the Hubble parameter tension}",
  journal = "Phys. Dark Univ.",
  volume = "30",
  pages = "100662",
  year = "2020",
  doi = "10.1016/j.dark.2020.100662",
  eprint = "2004.13163",
  archivePrefix = "arXiv",
  primaryClass = "gr-qc"
}

@article{MartinsVieira2024,
  author = "Martins, C. J. A. P. and Vieira, A. M. M.",
  title = "{Current and future cosmological impact of microwave background temperature measurements}",
  journal = "Phys. Dark Univ.",
  volume = "44",
  pages = "101494",
  year = "2024",
  doi = "10.1016/j.dark.2024.101494",
  eprint = "2404.05037",
  archivePrefix = "arXiv",
  primaryClass = "astro-ph.CO"
}

@article{nogo,
  author = "Izaurieta, F. and Lepe, S. and Quinzacara, C.",
  title = "{No late-time role for adiabatic torsion: a no-go result for Hubble-cutoff holographic dark energy in Einstein--Cartan cosmology}",
  journal = "Eur. Phys. J. C",
  volume = "86",
  pages = "1088",
  year = "2026",
  doi = "10.1140/epjc/s10052-026-16337-0",
  eprint = "2607.16370",
  archivePrefix = "arXiv",
  primaryClass = "gr-qc"
}

@article{Izaurieta:2019dix,
  author = "Izaurieta, F. and Rodr{\'\i}guez, E. and Valdivia, O.",
  title = "{Linear and second-order geometry perturbations on spacetimes with torsion}",
  journal = "Eur. Phys. J. C",
  volume = "79",
  pages = "337",
  year = "2019",
  doi = "10.1140/epjc/s10052-019-6852-y",
  eprint = "1901.06400",
  archivePrefix = "arXiv",
  primaryClass = "gr-qc"
}

@article{Barrientos:2019msu,
  author = "Barrientos, J. and Izaurieta, F. and Rodr{\'\i}guez, E. and Valdivia, O.",
  title = "{Wave operators, torsion, and Weitzenb{\"o}ck identities}",
  journal = "Gen. Relativ. Gravit.",
  volume = "54",
  pages = "26",
  year = "2022",
  doi = "10.1007/s10714-022-02914-7",
  eprint = "1903.04712",
  archivePrefix = "arXiv",
  primaryClass = "gr-qc"
}

@article{Shapiro2002,
  author = "Shapiro, I. L.",
  title = "{Physical aspects of the space-time torsion}",
  journal = "Phys. Rept.",
  volume = "357",
  pages = "113",
  year = "2002",
  doi = "10.1016/S0370-1573(01)00030-8",
  eprint = "hep-th/0103093",
  archivePrefix = "arXiv",
  primaryClass = "hep-th"
}

@misc{MMN2026,
  author = "Ma, X. and Mudrunka, K. and Nakayama, K.",
  title = "{Inflation and reheating by dynamical torsion}",
  year = "2026",
  eprint = "2607.29618",
  archivePrefix = "arXiv",
  primaryClass = "hep-ph"
}

@article{Capolupo2026,
  author = "Capolupo, A. and Carloni, S. and Fabbri, L. and Monda, S. and Quaranta, A. and Vignolo, S.",
  title = "{Quantized Dirac fields in torsionful gravity: cosmological implications and links with the dark universe}",
  journal = "Int. J. Geom. Meth. Mod. Phys.",
  volume = "23",
  pages = "2650012",
  year = "2026",
  doi = "10.1142/S021988782650012X",
  eprint = "2510.06874",
  archivePrefix = "arXiv",
  primaryClass = "hep-th"
}

@article{HHK1974,
  author = "Hehl, F. W. and von der Heyde, P. and Kerlick, G. D.",
  title = "{General relativity with spin and torsion and its deviations from Einstein's theory}",
  journal = "Phys. Rev. D",
  volume = "10",
  pages = "1066",
  year = "1974",
  doi = "10.1103/PhysRevD.10.1066"
}

@article{Poplawski2012,
  author = "Pop{\l}awski, N. J.",
  title = "{Nonsingular, big-bounce cosmology from spinor-torsion coupling}",
  journal = "Phys. Rev. D",
  volume = "85",
  pages = "107502",
  year = "2012",
  doi = "10.1103/PhysRevD.85.107502",
  eprint = "1111.4595",
  archivePrefix = "arXiv"
}

@article{FreidelMinicTakeuchi2005,
  author = "Freidel, L. and Minic, D. and Takeuchi, T.",
  title = "{Quantum gravity, torsion, parity violation, and all that}",
  journal = "Phys. Rev. D",
  volume = "72",
  pages = "104002",
  year = "2005",
  doi = "10.1103/PhysRevD.72.104002",
  eprint = "hep-th/0507253",
  archivePrefix = "arXiv",
  primaryClass = "hep-th"
}

@article{MZK2013,
  author = "Magueijo, J. and Z{\l}o{\'s}nik, T. G. and Kibble, T. W. B.",
  title = "{Cosmology with a spin}",
  journal = "Phys. Rev. D",
  volume = "87",
  pages = "063504",
  year = "2013",
  doi = "10.1103/PhysRevD.87.063504",
  eprint = "1212.0585",
  archivePrefix = "arXiv",
  primaryClass = "astro-ph.CO"
}

@article{Hojman1978,
  author = "Hojman, S. and Rosenbaum, M. and Ryan, M. P. and Shepley, L. C.",
  title = "{Gauge invariance, minimal coupling, and torsion}",
  journal = "Phys. Rev. D",
  volume = "17",
  pages = "3141",
  year = "1978",
  doi = "10.1103/PhysRevD.17.3141"
}

@article{KasemKhalil2021,
  author = "Kasem, A. and Khalil, S.",
  title = "{Gauge-field-induced torsion and cosmic inflation}",
  journal = "Int. J. Mod. Phys. A",
  volume = "36",
  pages = "2150161",
  year = "2021",
  doi = "10.1142/S0217751X2150161X",
  eprint = "2005.06228",
  archivePrefix = "arXiv",
  primaryClass = "hep-th"
}

@article{Bolejko2020,
  author = "Bolejko, K. and Cinus, M. and Roukema, B. F.",
  title = "{Cosmological signatures of torsion and how to distinguish torsion from the dark sector}",
  journal = "Phys. Rev. D",
  volume = "101",
  pages = "104046",
  year = "2020",
  doi = "10.1103/PhysRevD.101.104046",
  eprint = "2003.06528",
  archivePrefix = "arXiv",
  primaryClass = "astro-ph.CO"
}

@article{Tsamparlis1979,
  author = "Tsamparlis, M.",
  title = "{Cosmological principle and torsion}",
  journal = "Phys. Lett. A",
  volume = "75",
  pages = "27",
  year = "1979",
  doi = "10.1016/0375-9601(79)90265-2"
}

@article{Cid2018,
  author = "Cid, A. and Izaurieta, F. and Leon, G. and Medina, P. and Narbona, D.",
  title = "{Non-minimally coupled scalar field cosmology with torsion}",
  journal = "JCAP",
  volume = "04",
  pages = "041",
  year = "2018",
  doi = "10.1088/1475-7516/2018/04/041",
  eprint = "1704.04563",
  archivePrefix = "arXiv",
  primaryClass = "gr-qc"
}

@article{PTB2017,
  author = "Pasmatsiou, K. and Tsagas, C. G. and Barrow, J. D.",
  title = "{Kinematics of Einstein--Cartan universes}",
  journal = "Phys. Rev. D",
  volume = "95",
  pages = "104007",
  year = "2017",
  doi = "10.1103/PhysRevD.95.104007",
  eprint = "1611.07878",
  archivePrefix = "arXiv",
  primaryClass = "gr-qc"
}

@article{ITP2018,
  author = "Iosifidis, D. and Tsagas, C. G. and Petkou, A. C.",
  title = "{The Raychaudhuri equation in spacetimes with torsion and non-metricity}",
  journal = "Phys. Rev. D",
  volume = "98",
  pages = "104037",
  year = "2018",
  doi = "10.1103/PhysRevD.98.104037",
  eprint = "1809.04992",
  archivePrefix = "arXiv",
  primaryClass = "gr-qc"
}

@article{ObukhovTresguerres1993,
  author = "Obukhov, {\relax Yu}. N. and Tresguerres, R.",
  title = "{Hyperfluid---a model of classical matter with hypermomentum}",
  journal = "Phys. Lett. A",
  volume = "184",
  pages = "17",
  year = "1993",
  doi = "10.1016/0375-9601(93)90339-2",
  eprint = "gr-qc/0008013",
  archivePrefix = "arXiv",
  primaryClass = "gr-qc"
}

@article{ChlubaSunyaev2012,
  author = "Chluba, J. and Sunyaev, R. A.",
  title = "{The evolution of CMB spectral distortions in the early Universe}",
  journal = "Mon. Not. R. Astron. Soc.",
  volume = "419",
  pages = "1294",
  year = "2012",
  doi = "10.1111/j.1365-2966.2011.19786.x",
  eprint = "1109.6552",
  archivePrefix = "arXiv",
  primaryClass = "astro-ph.CO"
}

@article{FIRAS1996,
  author = "Fixsen, D. J. and Cheng, E. S. and Gales, J. M. and Mather, J. C. and Shafer, R. A. and Wright, E. L.",
  title = "{The cosmic microwave background spectrum from the full COBE FIRAS data set}",
  journal = "Astrophys. J.",
  volume = "473",
  pages = "576",
  year = "1996",
  doi = "10.1086/178173",
  eprint = "astro-ph/9605054",
  archivePrefix = "arXiv",
  primaryClass = "astro-ph"
}

@article{Chluba2014,
  author = "Chluba, J.",
  title = "{Tests of the CMB temperature--redshift relation, CMB spectral distortions and why adiabatic photon production is hard}",
  journal = "Mon. Not. R. Astron. Soc.",
  volume = "443",
  pages = "1881",
  year = "2014",
  doi = "10.1093/mnras/stu1260",
  eprint = "1405.1277",
  archivePrefix = "arXiv",
  primaryClass = "astro-ph.CO"
}

@article{Lima1996,
  author = "Lima, J. A. S.",
  title = "{Thermodynamics of decaying vacuum cosmologies}",
  journal = "Phys. Rev. D",
  volume = "54",
  pages = "2571",
  year = "1996",
  doi = "10.1103/PhysRevD.54.2571",
  eprint = "gr-qc/9605055",
  archivePrefix = "arXiv",
  primaryClass = "gr-qc"
}

@article{LSV2000,
  author = "Lima, J. A. S. and Silva, A. I. and Viegas, S. M.",
  title = "{Is the radiation temperature--redshift relation of the standard cosmology in accordance with the data?}",
  journal = "Mon. Not. R. Astron. Soc.",
  volume = "312",
  pages = "747",
  year = "2000"
}

@article{Jetzer2011,
  author = "Jetzer, P. and Puy, D. and Signore, M. and Tortora, C.",
  title = "{Limits on decaying dark energy density models from the CMB temperature--redshift relation}",
  journal = "Gen. Relativ. Gravit.",
  volume = "43",
  pages = "1083",
  year = "2011",
  doi = "10.1007/s10714-010-1091-4",
  eprint = "1007.2325",
  archivePrefix = "arXiv",
  primaryClass = "astro-ph.CO"
}

@article{Kotani2026,
  author = "Kotani, T. and Oka, T. and Enokiya, R.",
  title = "{First determination of the cosmic microwave background radiation temperature at $z=0.68$ using molecular absorption lines}",
  journal = "Astrophys. J.",
  volume = "1000",
  pages = "141",
  year = "2026",
  doi = "10.3847/1538-4357/ae4aa0",
  eprint = "2602.22399",
  archivePrefix = "arXiv",
  primaryClass = "astro-ph.CO"
}

@article{Avgoustidis2016,
  author = "Avgoustidis, A. and G{\'e}nova-Santos, R. T. and Luzzi, G. and Martins, C. J. A. P.",
  title = "{Subpercent constraints on the cosmological temperature evolution}",
  journal = "Phys. Rev. D",
  volume = "93",
  pages = "043521",
  year = "2016",
  doi = "10.1103/PhysRevD.93.043521",
  eprint = "1511.04335",
  archivePrefix = "arXiv",
  primaryClass = "astro-ph.CO"
}

@article{Planck2018,
  author = "{Planck Collaboration} and Aghanim, N. and others",
  title = "{Planck 2018 results. VI. Cosmological parameters}",
  journal = "Astron. Astrophys.",
  volume = "641",
  pages = "A6",
  year = "2020",
  doi = "10.1051/0004-6361/201833910",
  eprint = "1807.06209",
  archivePrefix = "arXiv",
  primaryClass = "astro-ph.CO"
}

@article{Planck2018III,
  author = "{Planck Collaboration} and Aghanim, N. and others",
  title = "{Planck 2018 results. III. High Frequency Instrument data processing and frequency maps}",
  journal = "Astron. Astrophys.",
  volume = "641",
  pages = "A3",
  year = "2020",
  doi = "10.1051/0004-6361/201832909",
  eprint = "1807.06207",
  archivePrefix = "arXiv",
  primaryClass = "astro-ph.CO"
}

@article{Planck2018V,
  author = "{Planck Collaboration} and Aghanim, N. and others",
  title = "{Planck 2018 results. V. CMB power spectra and likelihoods}",
  journal = "Astron. Astrophys.",
  volume = "641",
  pages = "A5",
  year = "2020",
  doi = "10.1051/0004-6361/201936386",
  eprint = "1907.12875",
  archivePrefix = "arXiv",
  primaryClass = "astro-ph.CO"
}

@article{PLJH2019,
  author = "Pereira, S. H. and Lima, {\relax R. de C} and Jesus, J. F. and Holanda, R. F. L.",
  title = "{Acceleration in Friedmann cosmology with torsion}",
  journal = "Eur. Phys. J. C",
  volume = "79",
  pages = "950",
  year = "2019",
  doi = "10.1140/epjc/s10052-019-7462-4",
  eprint = "1906.07624",
  archivePrefix = "arXiv",
  primaryClass = "gr-qc"
}

@article{VJP2025,
  author = "Vicente, A. M. and Jesus, J. F. and Pereira, S. H.",
  title = "{Kinematic reconstruction of torsion as dark energy in Friedmann cosmology}",
  journal = "JCAP",
  volume = "03",
  pages = "034",
  year = "2025",
  doi = "10.1088/1475-7516/2025/03/034",
  eprint = "2402.00844",
  archivePrefix = "arXiv",
  primaryClass = "gr-qc"
}

@article{Liu2025,
  author = "Liu, T. and Li, X. and Xu, T. and Biesiada, M. and Wang, J.",
  title = "{Torsion cosmology in the light of DESI, supernovae and CMB observational constraints}",
  journal = "Eur. Phys. J. C",
  volume = "85",
  pages = "1351",
  year = "2025",
  doi = "10.1140/epjc/s10052-025-15090-0",
  eprint = "2507.04265",
  archivePrefix = "arXiv",
  primaryClass = "astro-ph.CO"
}

@article{CsillagHarko2024,
  author = "Csillag, L. and Harko, T.",
  title = "{Semi-symmetric metric gravity: From the Friedmann--Schouten geometry with torsion to dynamical dark energy models}",
  journal = "Phys. Dark Univ.",
  volume = "46",
  pages = "101596",
  year = "2024",
  doi = "10.1016/j.dark.2024.101596",
  eprint = "2402.06114",
  archivePrefix = "arXiv",
  primaryClass = "gr-qc"
}

@misc{CapolupoZeldovich2026,
  author = "Capolupo, A. and Monda, S. and Pisacane, G. and Quaranta, A. and Serao, R.",
  title = "{Possible microscopic source of torsion}",
  year = "2026",
  eprint = "2610.28125",
  archivePrefix = "arXiv",
  primaryClass = "hep-th",
  note = "Contribution to the Sixth Zeldovich Meeting (Pescara, 2026), submitted to Astron. Rep."
}

@article{TilquinSchucker2011,
  author = "Tilquin, A. and Sch{\"u}cker, T.",
  title = "{Torsion, an alternative to dark matter?}",
  journal = "Gen. Relativ. Gravit.",
  volume = "43",
  pages = "2965",
  year = "2011",
  doi = "10.1007/s10714-011-1222-6",
  eprint = "1104.0160",
  archivePrefix = "arXiv",
  primaryClass = "astro-ph.CO"
}

@misc{WhoFeeds,
  author = "Izaurieta, F. and Lepe, S. and Quinzacara, C.",
  title = "{Who feeds the torsion? Spin density, dilution and the source of torsion dark energy}",
  year = "2026",
  note = "In preparation"
}

@misc{PaperB,
  author = "Izaurieta, F. and Lepe, S. and Quinzacara, C.",
  title = "{Torsion accelerates but does not inflate: a pseudo-inflation theorem for Einstein--Cartan cosmology}",
  year = "2026",
  note = "Companion paper, submitted to Phys.\ Rev.\ D"
}

@article{ECportal2021,
  author = "Shaposhnikov, M. and Shkerin, A. and Timiryasov, I. and Zell, S.",
  title = "{Einstein--Cartan portal to dark matter}",
  journal = "Phys. Rev. Lett.",
  volume = "126",
  pages = "161301",
  year = "2021",
  doi = "10.1103/PhysRevLett.126.161301",
  eprint = "2008.11686",
  archivePrefix = "arXiv",
  primaryClass = "hep-ph"
}

@book{Hodge1941,
  author = "Hodge, W. V. D.",
  title = "{The Theory and Applications of Harmonic Integrals}",
  publisher = "Cambridge University Press",
  address = "Cambridge",
  year = "1941"
}

@book{Flanders1963,
  author = "Flanders, H.",
  title = "{Differential Forms with Applications to the Physical Sciences}",
  publisher = "Academic Press",
  address = "New York",
  year = "1963"
}

@article{Ellis2013,
  author = "Ellis, G. F. R. and Poltis, R. and Uzan, J.-P. and Weltman, A.",
  title = "{Blackness of the cosmic microwave background spectrum as a probe of the distance-duality relation}",
  journal = "Phys. Rev. D",
  volume = "87",
  pages = "103530",
  year = "2013",
  doi = "10.1103/PhysRevD.87.103530",
  eprint = "1301.1312",
  archivePrefix = "arXiv",
  primaryClass = "astro-ph.CO"
}

@article{ARCADE2,
  author = "Fixsen, D. J. and Kogut, A. and Levin, S. and Limon, M. and Lubin, P. and Mirel, P. and Seiffert, M. and Singal, J. and Wollack, E. and Villela, T. and Wuensche, C. A.",
  title = "{ARCADE 2 measurement of the absolute sky brightness at 3--90 GHz}",
  journal = "Astrophys. J.",
  volume = "734",
  pages = "5",
  year = "2011",
  doi = "10.1088/0004-637X/734/1/5",
  eprint = "0901.0555",
  archivePrefix = "arXiv",
  primaryClass = "astro-ph.CO"
}

@article{Avgoustidis2012,
  author = "Avgoustidis, A. and Luzzi, G. and Martins, C. J. A. P. and Monteiro, A. M. R. V. L.",
  title = "{Constraints on the CMB temperature--redshift dependence from SZ and distance measurements}",
  journal = "JCAP",
  volume = "02",
  pages = "013",
  year = "2012",
  doi = "10.1088/1475-7516/2012/02/013",
  eprint = "1112.1862",
  archivePrefix = "arXiv",
  primaryClass = "astro-ph.CO"
}

@article{Uzan2011,
  author = "Uzan, J.-P.",
  title = "{Varying constants, gravitation and cosmology}",
  journal = "Living Rev. Relativ.",
  volume = "14",
  pages = "2",
  year = "2011",
  doi = "10.12942/lrr-2011-2",
  eprint = "1009.5514",
  archivePrefix = "arXiv",
  primaryClass = "astro-ph.CO"
}

\end{document}